\documentclass[reprint,nofootinbib,amsmath,amssymb,aps,floatfix,superscriptaddress,twocolumn]{revtex4-2}
\usepackage{graphicx}
\usepackage{dcolumn}
\usepackage[colorlinks,linkcolor=magenta,anchorcolor=cyan,citecolor=blue]{hyperref}
\usepackage{bm}
\usepackage[multiple]{footmisc}

\begin{document}
\title{Observational Signatures of Thin Accretion Disks around Rotating Black Holes Embedded in Dark Matter Halos}

\author{Jia-Ying Zhang}  
\author{Wei-Lun Tong}
\author{Tian-Yun Ye}
\author{Yong-You Liu}

\author{Zhen Li} 
\email{zhen.li@just.edu.cn}
\affiliation{School of Science, Jiangsu University of Science and Technology, Zhenjiang 212100, China}

\date{\today}

\begin{abstract}
Black holes embedded in a dark matter background represent an important candidate model that extends the standard vacuum case, characterizing the strong gravitational environments at galactic centers and  the gravitational influence of dark matter. This study focuses on dark matter halo-modified Kerr black holes and systematically investigates the observational characteristics of their surrounding thin accretion disks. Starting from static black hole solutions with dark matter halos, we employ the Newman--Janis algorithm to construct its rotating counterpart. The dark matter halo is modeled by a general double power-law density profile, whose integrated mass $M_D(r)$ modifies the radial mass function of Kerr metric to $m(r) = M + M_D(r)$. Based on this spacetime, we analyze the observational properties of their thin accretion disk, including the radiative flux, temperature, differential luminosity, and spectral luminosity as a function of radius, with respect to the black hole spin and dark matter halo parameters. Furthermore, by combining ray-tracing methods, we simulate the bolometric image of the thin accretion disk under different black hole spins, dark matter halo parameters, and viewing inclinations. The dark matter halo enhances the gravitational potential, shifting the event horizon, innermost stable circular orbit, and photon shell outward, suppressing the peak radiative flux, and producing a fainter image, thereby generating signatures that are possibly observational distinguishable from standard vacuum Kerr black holes.
\end{abstract}
\maketitle

\section{Introduction}

As the most gravitationally extreme compact objects predicted by general relativity, black holes serve as ideal natural laboratories for testing gravitational theories and probing exotic physical processes within strong-field spacetime regimes. Driven by cutting-edge astronomical observations, black hole astrophysics has witnessed rapid advances in recent years. The Event Horizon Telescope (EHT) has acquired high-resolution shadow images of M87* (Messier 87*) \cite{EHT2019M87} and Sgr A* (Sagittarius A*) \cite{EHT2022SgrA}, which are fully compatible with the expectations based on the existence of black hole horizons. Meanwhile, multi-band spectroscopic and photometric data of accreting black holes have been continuously accumulated from active galactic nuclei \cite{Kang2025,Casura2024,Arevalo-Gonzalez2025,Juodzbalis2026} and X-ray binaries \cite{Tremou2026,John2024,Tetarenko2016,Buisson2019}. As the canonical framework linking black hole intrinsic spacetime geometry to observable radiative signatures, the thin accretion disk paradigm serves as the primary framework for interpreting accretion-related astronomical measurements \cite{Novikov1973,Page1974,Shakura1973,Hankla2025,Hagen2024,Luminet1979}. Consequently, theoretical modeling of black hole spacetime, accretion structures, and shadow characteristics has become a central research avenue in modern gravitational astrophysics \cite{Hu2025,Kumar2020,Li2021,Chael2021,Santibanez2025,Chang2025,Hou2022,Feng2024,Guerrero2021,Liu2022,BisnovatyiKogan2022,Afrin2023, guo1,guo2,guo3,yong1,yong2,wang}. 

Nevertheless, the majority of existing theoretical studies rely on the vacuum Kerr black hole \cite{Kerr1963,Johnson1,Gralla1}, which assumes an isolated spacetime environment and neglects gravitational perturbations from surrounding galactic matter components. Furthermore, cosmological observations have robustly confirmed that dark matter dominates the total mass budget of nearly all galaxies \cite{deGraaff2024,Shankar2025,Sarkar2026}, such that every supermassive black hole at the galactic center is naturally immersed in a dense dark matter halo. This realistic cosmic environment exposes two notable limitations in current mainstream theoretical frameworks. First, the vacuum Kerr metric cannot characterize spacetime geometric corrections induced by dark matter gravity \cite{Liu2023}. Consequently, theoretical results of accretion disk morphology, orbital kinematics and photon propagation contain significant systematic biases compared with real astrophysical scenarios. Second, the limited existing research on dark matter-embedded rotating black holes typically examine individual dark matter halo density profiles, lacking generality \cite{Hou2018,Shen2024,Mora2025,Zhu2019}. These deficiencies impede the application of existing theoretical frameworks to the interpretation of high-precision EHT and X-ray survey observational data.

Against this research background, the present work focuses on a self-consistent theoretical model for rotating black holes within general double power-law dark matter halos \cite{37}. We first derive the dark matter halo-modified rotating spacetime metric by employing the Newman--Janis algorithm \cite{nj1,nj2}, which provides a well-established procedure for generating rotating solutions from static and spherically symmetric seed metrics. Then we solve the timelike and null geodesic equations of test particles in this spacetime. Based on the derived orbital and spacetime quantities, we derive analytical expressions for disk radiative flux, surface temperature, differential luminosity and spectral luminosity distribution within the framework of standard thin accretion disk model, and quantitatively characterize the combined modulation effect of black hole spin and dark matter halo parameters. To further investigate the observational manifestations, we adopt the ray-tracing numerical algorithm to simulate the bolometric image of dark matter halo-modified Kerr black hole, with thin accretion disk as the emission source. Through comparative analysis with standard Kerr black hole, we elaborate how dark matter halo reshapes the observed images.

The remainder of this paper is organized as follows. In Sec.~\ref{sec2}, we construct the rotating black hole metric with dark matter halos, using the Newman-Janis method. In Sec.~\ref{sec3}, we analyze the timelike and null geodesic equations in this spacetime. In Sec.~\ref{sec4}, we investigate the radiative flux, temperature profile, differential and spectral luminosity of the thin accretion disk. In Sec.~\ref{sec5}, we employ the backward ray-tracing algorithm to simulate the black hole image, and discuss the corresponding observational discriminants. Finally, Sec.~\ref{sec6} summarizes our main results and presents future prospects. Geometric units are consistently adopted throughout this work, with the gravitational constant $G$ and the speed of light $c$ both set to unity.

\section{Rotating black holes with dark matter halos}\label{sec2}

We start from the static, spherically symmetric metric with dark matter halo. It is obtained by using the Zhao's dark matter density profile, which is given by \cite{37}
\begin{equation}\label{za}
\rho(r) = \rho_s \left( \frac{r}{r_s} \right)^{-\gamma} \left[ \left( \frac{r}{r_s} \right)^{\alpha} + 1 \right]^{(\gamma - \beta)/\alpha},
\end{equation}
where $\rho_s$ and $r_s$ denote the characteristic density and scale radius, respectively, while $\alpha$, $\beta$, and $\gamma$ control the shape of the profile. Specifically, $\gamma$ and $\beta$ respectively set the inner and outer slopes, and $\alpha$ determines the sharpness of the transition between these two regimes. All five parameters are nonnegative. The Zhao's density profile provides a unified parametrization capable of reproducing a wide range of commonly used dark matter distributions through different choices of its characteristic parameters, including the Navarro-Frenk-White (NFW) profile ($\alpha,\beta,\gamma=1,3,1$) \cite{nfw}, the Hernquist profile \cite{heq} ($\alpha,\beta,\gamma=1,4,1$), and many of their profiles \cite{prof1,prof2,prof3,prof4,prof5,prof7,prof8,prof9}. Consequently, the rotating solution derived in this work is not merely associated with a single halo model, but constitutes a general framework from which many existing models can be recovered as special cases. This unified treatment enables a systematic exploration of how different halo model, ranging from cored to cuspy inner structures and from shallow to steep outer density profiles, affect the spacetime geometry and the corresponding optical observables.

The Zhao's profile leads to the static and spherically symmetric metric as follow \cite{zli}
\begin{equation}\label{static}
ds^2 = -f(r) dt^2 + \frac{dr^2}{f(r)} + h(r) \left( d\theta^2 + \sin^2\theta \, d\phi^2 \right),
\end{equation}
with $h(r)=r^2$ and
\begin{align}
f(r) &= 1 - \frac{2m(r)}{r},\\
m(r) &= M + M_D(r)\\
M_D(r) &= \frac{4\pi\rho_s r_s^3}{\alpha}\, B\!\left( \frac{(r/r_s)^\alpha}{1 + (r/r_s)^\alpha};\, \frac{3-\gamma}{\alpha},\, \frac{\beta-3}{\alpha} \right).\label{mr}
\end{align}
Here $B(z; p,q)$ is the incomplete beta function, and $M$ is the central black hole mass.

\subsection{The Newman-Janis method}

To obtain its rotating counterpart, we apply the Newman-Janis algorithm \cite{nj1,nj2}. First, we transform the coordinates $(t,r,\theta,\phi)$ in Eq.~(\ref{static}) to Eddington-Finkelstein coordinates $(u,r,\theta,\phi)$ via
\begin{equation}
du = dt - \frac{dr}{f(r)}.
\end{equation}
In these coordinates the inverse metric can be expressed in terms of a null tetrad $(l^\mu, n^\mu, m^\mu, \bar m^\mu)$ as
\begin{equation}\label{tet}
g^{\mu\nu} = -l^\mu n^\nu - l^\nu n^\mu + m^\mu \bar m^\nu + m^\nu \bar m^\mu,
\end{equation}
with the tetrad vectors satisfying
\begin{align}
l^\mu l_\mu & = n^\mu n_\mu = m^\mu m_\mu = \bar m^\mu \bar m_\mu =0, \\
l^\mu m_\mu &= l^\mu \bar m_\mu = n^\mu m_\mu = n^\mu \bar m_\mu =0,\\
l^\mu n_\mu &= -m^\mu \bar m_\mu = -1.
\end{align}
A convenient choice is \cite{nj3,nj4}
\begin{align}
l^\mu &= \delta_r^\mu,\\
n^\mu &= \delta_u^\mu - \frac{f(r)}{2} \delta_r^\mu,\\
m^\mu &= \frac{1}{\sqrt{2h(r)}} \left( \delta_\theta^\mu + \frac{i}{\sin\theta} \delta_\phi^\mu \right),\\
\bar m^\mu &= \frac{1}{\sqrt{2h(r)}} \left( \delta_\theta^\mu - \frac{i}{\sin\theta} \delta_\phi^\mu \right).
\end{align}

The Newman-Janis method proceeds by performing a complex transformation in the $r$–$u$ plane:
\begin{equation}
r \to r' = r + i a \cos\theta, \qquad u \to u' = u - i a \cos\theta,
\end{equation}
where $a$ is the black hole spin, and leave $\theta$ and $\phi$ unchanged. After these transformations, the metric function also takes a new form 
\begin{equation}
f(r)\to \tilde f(r'),\quad h(r)\to \tilde h(r').
\end{equation}
A comprehensive discussion of the complexification procedure can be found in \cite{nj5}. In order to recover the Kerr solution in the limit of vanishing dark matter, we adopt the following transformation form of metric functions \cite{nj3,nj4}
\begin{align}
\tilde h(r') &= \Sigma \equiv r^2 + a^2\cos^2\theta,\label{ht}\\
\tilde f(r') &= 1 - \frac{2m(r) r}{\Sigma},\label{ft}
\end{align}
where the mass function $m(r)$ is assumed to remain a function of the real radial coordinate $r$ only. The tetrads are then transformed to \cite{nj3,nj4}
\begin{align}
l^{\prime\mu} &= \delta_r^\mu,\\
n^{\prime\mu} &= \delta_u^\mu - \frac{\tilde f(r')}{2} \delta_r^\mu,\\
m^{\prime\mu} &= \frac{1}{\sqrt{2\tilde h(r')}} \left( i a \sin\theta\,(\delta_u^\mu - \delta_r^\mu) + \delta_\theta^\mu + \frac{i}{\sin\theta} \delta_\phi^\mu \right),\\
\bar m^{\prime\mu} &= \frac{1}{\sqrt{2\tilde h(r')}} \left( -i a \sin\theta\,(\delta_u^\mu - \delta_r^\mu) + \delta_\theta^\mu - \frac{i}{\sin\theta} \delta_\phi^\mu \right).
\end{align}
Using this new tetrad, based on Eq.~(\ref{tet}), we obtain the metric in null coordinates:
\begin{align}
g_{uu} &= - \tilde f(r'), \quad g_{ur}=g_{ru}=-1,\\
g_{u\phi} &= g_{\phi u} = a\sin^2\theta\bigl(\tilde f(r')-1\bigr),\\
g_{r\phi} &= g_{\phi r} =  a\sin^2\theta,\\
g_{\theta\theta} &= \tilde h(r'),\\
g_{\phi\phi} &=\sin^2\theta\Bigl[\tilde h(r') + a^2\sin^2\theta\bigl(2-\tilde f(r')\bigr)\Bigr].
\end{align}

To express the metric in the familiar Boyer–Lindquist form, we perform the coordinate transformation
\begin{equation}
du = dt + F(r)\,dr,\qquad d\phi  = d\varphi + G(r)\,dr,
\end{equation}
where $F(r)$ and $G(r)$ can be determined by imposing the condition that all off-diagonal components of the metric tensor vanish except for the $t\varphi$ and $\varphi t$ components. Following Azreg-A{\"\i}nou prescription, they are given by \cite{nj3,nj4}
\begin{align}
F(r) &= - \frac{h(r) + a^2}
{f(r)h(r) + a^2},\\
G(r) &= - \frac{a}
{f(r)h(r) + a^2}.
\end{align}
After coordinate transformation, the non-vanishing metric components become
\begin{equation}
\begin{aligned}
g_{tt} &= - \tilde f(r'),\\
g_{t\varphi} &= g_{\varphi t} = a\sin^2\theta\bigl(\tilde f(r')-1\bigr),\\
g_{rr} &= \frac{\tilde h(r')}
{h(r)f(r) + a^2},\\
g_{\theta\theta} &= \tilde h(r'),\\
g_{\varphi \varphi} &= \sin^2\theta\Bigl[\tilde h(r') + a^2\sin^2\theta\bigl(2-\tilde f(r')\bigr)\Bigr].
\end{aligned}
\label{eq:metric_part}
\end{equation}


After substituting Eq.(\ref{ht}) and (\ref{ft}), the line element in Boyer–Lindquist coordinates can be written as
\begin{equation}\label{metric}
ds^2 = g_{tt}\,dt^2 + g_{rr}\,dr^2 + g_{\theta\theta}\,d\theta^2 + g_{\varphi\varphi}\,d\varphi^2 + 2g_{t\varphi}\,dt\,d\varphi,
\end{equation}
where 
\begin{align}
g_{tt} &= -\left[\frac{\Delta - a^2\sin^2\theta}{\Sigma}\right],\\
g_{rr} &= \frac{\Sigma}{\Delta},\\
g_{t\varphi} &= -a\sin^2\theta\left[1 - \frac{\Delta - a^2\sin^2\theta}{\Sigma}\right],\\
g_{\theta\theta} &= \Sigma,\\
g_{\varphi \varphi} &= \sin^2\theta\left[\Sigma + a^2\sin^2\theta\left(2 - \frac{\Delta - a^2\sin^2\theta}{\Sigma}\right)\right],
\end{align}
and we have introduced
\begin{equation}
\Delta = r^2 + a^2 - 2m(r)r, \quad\Sigma = r^2 + a^2\cos^2\theta.
\end{equation}
The radially dependent mass function $m(r)$ characterizes the spacetime of rotating black holes surrounded by dark matter halos. When $M_D(r) = 0$, the model reduces to the vacuum Kerr metric, which describes an isolated rotating black hole in a dark matter free environment.

\begin{figure}
    \centering
    \includegraphics[scale=0.55]{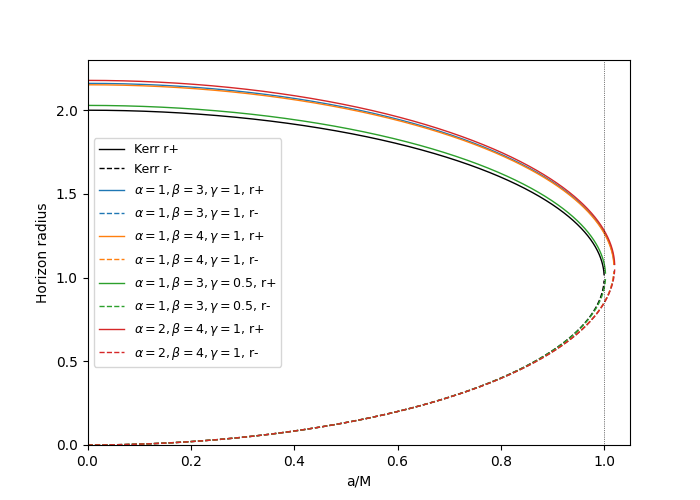}
    \caption{The horizons $r_\pm$ as a function of black hole spin $a/M$, with different combinations of halo parameters $\alpha,\beta,\gamma$=(1,3,1), (1,4,1), (1,3,0.5), (2,4,1), while we set $\rho_sM^2=0.0001$ and $r_s/M=30$. The Kerr case is also shown as a reference. The solid and dashed curves represent the outer event horizon and inner horizon, respectively.}
\label{horz}
\end{figure}

The locations of the horizons are determined by the roots of
\begin{equation}
\Delta(r)=r^2 + a^2 - 2m(r)r=0.
\end{equation}
To investigate the horizon structure more systematically, we solve the equation $\Delta(r)=0$ numerically and determine the inner horizon $r_-$ and outer horizon $r_+$ as functions of the spin parameter. The results are shown in Fig.~\ref{horz}. Several interesting trends emerge. First, the outer horizon is significantly more sensitive to the dark matter halo than the inner horizon. Physically, the outer horizon is located farther away from the black hole center and therefore encloses a larger amount of dark matter mass. Since the halo contribution accumulates with radius, its effect naturally becomes more pronounced at the outer horizon. By contrast, the inner horizon resides deep inside the strong-field region where the enclosed dark matter mass is comparatively small, resulting in only minor deviations from the Kerr prediction.

Second, the presence of dark matter shifts the extremal limit of the rotating black hole. In the Kerr spacetime, extremality occurs at $a/M=1$, where the two horizons merge into a single degenerate horizon. In the present model, however, the additional gravitational mass contributed by the dark matter halo modifies this condition. As shown in Fig.~\ref{horz}, the horizon structure can remain even for spin parameters slightly exceeding $a/M=1$. This does not imply a violation of cosmic censorship; rather, it reflects the fact that the total gravitating mass of the system is larger than the bare black hole mass $M$ due to the surrounding dark matter distribution. Consequently, the effective extremal spin parameter can exceed unity when expressed in terms of the central black hole mass alone.

\subsection{Stress-energy tensor of dark matter halos}
The metric (\ref{metric}) obtained through the Newman–Janis procedure should be regarded as an effective rotating extension of the static dark matter black hole. The corresponding stress-energy tensor generally describes a rotating anisotropic matter distribution, and it can be computed directly from the Einstein field equations. In the following orthonormal tetrad~\cite{st},
\begin{align}
e_t^\mu &= -\frac{\left(r^2 + a^2,\, 0,\, 0,\, a\right)}{\sqrt{\Sigma\Delta}}, &
e_r^\mu &= -\frac{\sqrt{\Delta}\,(0,\, 1,\, 0,\, 0)}{\sqrt{\Sigma}}, \nonumber\\
e_\theta^\mu &= -\frac{(0,\, 0,\, 1,\, 0)}{\sqrt{\Sigma}}, &
e_\phi^\mu &= \frac{\left(a\sin^2\theta,\, 0,\, 0,\, 1\right)}{\sqrt{\Sigma}\sin\theta},
\end{align}
the Einstein tensor is diagonal, and hence the effective stress-energy tensor takes the form
\[
T_{\mu}^{\ \nu} = \operatorname{diag}\bigl(-\hat{\rho},\,\hat{p}_r,\,\hat{p}_\theta,\,\hat{p}_\phi\bigr),
\]
where $\hat{\rho}$ is the effective energy density, $\hat{p}_r$ the radial pressure, and $\hat{p}_\theta=\hat{p}_\phi$ the two transverse pressures. A straightforward calculation yields~\cite{st}
\begin{equation}
\hat{\rho} = -\hat{p}_r = \frac{rF_{,r} - F}{4\pi\Sigma^2},\qquad
\hat{p}_\theta = \hat{p}_\phi = -\hat{p}_r - \frac{F_{,rr}}{8\pi\Sigma},
\label{set}
\end{equation}
with $F(r)\equiv r m(r)$, and $F_{,r}\equiv dF/dr$. 

Using the explicit mass function $m(r)$ from Eq.~(\ref{mr}), we have $F_{,r} = m + r m'$ and $F_{,rr}=2m'+r m''$. Since $m'(r)=4\pi r^2\rho(r)$, the components in~(\ref{set}) reduce to
\begin{align}
\hat{\rho} &= \frac{r^4 \rho(r)}{\Sigma^2}, \label{rho}\\
\hat{p}_r &= -\frac{r^4 \rho(r)}{\Sigma^2}, \label{pr}\\
\hat{p}_\theta = \hat{p}_\phi &= \frac{r^4 \rho(r)}{\Sigma^2}
- \frac{4r^2\rho(r)+r^3\rho'(r)}{2\Sigma}. \label{pt}
\end{align}
These expressions reveal a clear physical structure demonstrating that the effective fluid is anisotropic.

In the static limit $a\to 0$, we have $\Sigma\to r^2$, and the above reduces to
\begin{align}
\hat{\rho} &= \rho(r),\qquad \hat{p}_r = -\rho(r),\\
\hat{p}_\theta &= \hat{p}_\phi = -\rho(r) - \frac{r}{2}\rho'(r),
\end{align}
which agrees exactly with the stress-energy tensor of the static spherically symmetric dark matter black hole derived in \cite{zli}. 

We now examine the energy conditions. In the orthonormal frame, we find $\hat\rho+\hat p_r =0$, so the radial null energy condition (NEC) is saturated. For the transverse directions, we assume a power-law halo density profile $\rho\propto r^{-n}$, where $n>0$ denotes the asymptotic index ($\alpha,\beta,\gamma$) of Eq.~(\ref{za}) valid in different radial regions. Then we have 
\begin{align}
\hat\rho + \hat p_\theta &= \frac{2r^4 \rho(r)}{\Sigma^2} -  \frac{4r^2\rho(r)+r^3\rho'(r)}{2\Sigma}\nonumber\\
&=\frac{r^2 \rho(r)}{2\Sigma^2} \big[n r^2 - (4-n) a^2 \cos^2\theta\big]
\end{align}

Since $\rho(r) > 0,\ \Sigma^2 > 0$, the necessary and sufficient condition for the transverse NEC to hold is:
\begin{equation}
n r^2 \ge (4-n) a^2 \cos^2\theta.
\end{equation}
Therefore, the transverse NEC is strictly satisfied everywhere at and outside the photon sphere. Violation of the inequality can occur only when $r$ lies very close to the horizon, the spin is near extremal, and the observer is located in the equatorial plane. This region lies well inside the photon sphere and therefore does not affect the subsequent shadow calculation.

\section{Geodesic properties}\label{sec3}

Owing to the stationarity and axisymmetry of the spacetime Eq.~(\ref{metric}), the metric components depend only on $r$ and $\theta$, and the geometry admits two Killing vectors $\partial_t$ and $\partial_\phi$, which give two conserved momentum components 
\begin{align}
p_t &= g_{tt} p^t + g_{t\phi} p^\phi = -E, \label{eq:6} \\
p_\phi &= g_{t\phi} p^t + g_{\phi\phi} p^\phi = L,\label{eq:7}
\end{align}
where $E$ and $L$ represent the particle energy and angular momentum. We also have the covariant momentum components, 
\begin{align}
  p^t = \frac{-E g_{\phi\phi} - L_z g_{t\phi}}{g_{tt}g_{\phi\phi} - g_{t\phi}^2}, \quad  
  p^\phi = \frac{L_z g_{tt} + E g_{t\phi}}{g_{tt}g_{\phi\phi} - g_{t\phi}^2}.
  \label{eq:eight}
\end{align}
The test particles in this modified spacetime have three conserved integrals of motion \cite{Bardeen1972}: the particle energy $E$, the axial angular momentum $L$, and the Carter constant $Q$. For simplicity, we introduce the dimensionless, energy-normalized parameters $\lambda \equiv L/E$ and $\eta \equiv Q/E^2$ to facilitate the analysis of the particle motion.

Substituting the metric components from Eq.~\eqref{eq:metric_part} and the normalized parameter $\lambda \equiv L/E$ into Eq.~\eqref{eq:eight}, we derive the geodesic equations governing temporal and azimuthal coordinate evolution via algebraic rearrangement:

\begin{align}
\frac{\Sigma}{E} p^\phi &= \frac{a}{\Delta} \left( r^2 + a^2 - a\lambda \right) + \frac{\lambda}{\sin^2\theta} - a,\label{eq:9} \\
\frac{\Sigma}{E} p^t &= \frac{r^2 + a^2}{\Delta} \left( r^2 + a^2 - a\lambda \right) + a \left( \lambda - a\sin^2\theta \right).\label{eq:10}
\end{align}

To analyze the radial and polar angular motion of test particles, we employ the Hamilton-Jacobi method, which is expressed as:

\begin{align}
   g^{\rho\nu} \frac{\partial S}{\partial x^\rho} \frac{\partial S}{\partial x^\nu} = -\mu^2,
\end{align}
where 
\begin{align}
    S = -Et + L\phi + S_r(r) + S_\theta(\theta).\,
\end{align}
is the action of test particles in separated variable form, satisfying \(p_\nu = \frac{\partial S}{\partial x^\nu}\).

The parameter $\mu$ denotes the rest mass of test particles: the limit $\mu = 0$ describes massless photons, while $\mu = 1$ is adopted for all massive particles. This normalization enables a unified treatment of massless and massive particle dynamics in terms of the conserved quantities $\lambda$ and $\eta$.

Substituting the separated-variable action \(S\) into the Hamilton–Jacobi equation and multiplying both sides by \(\Sigma\), we obtain the following equation:

\begin{align}
    \Delta \left( \frac{dS_r}{dr} \right)^2 + \left( \frac{dS_\theta}{d\theta} \right)^2 + g^{tt} E^2 + 2 g^{t\phi} E L + g^{\phi\phi} L^2 \nonumber\\= -\mu^2 \Sigma.
    \label{eq:13}
\end{align}

We can separate Eq.~\eqref{eq:13} into radial and angular components. First, we define the radial effective potential $\mathcal{R}(r)$ and polar angular effective potential $\Theta(\theta)$: 

\begin{align}
\mathcal{R}(r) &\equiv \left(r^2 + a^2 - a\lambda\right)^2 - \Delta\left[\mu^2 r^2 + \eta + (\lambda - a)^2\right], 
 \label{eq:14} \\
\Theta(\theta) &\equiv \eta + a^2 \cos^2\theta - \lambda^2 \cot^2\theta.
\end{align}
The geodesic equations are thus reduced to normalized radial and polar angular momentum components. 
\begin{align}
\frac{\Sigma}{E} p^r &= \pm_r \sqrt{\mathcal{R}(r)},\label{eq:16} \\
\frac{\Sigma}{E} p^\theta &= \pm_\theta \sqrt{\Theta(\theta)}.\label{eq:17} 
\end{align}
The sign $\pm_r$ distinguishes
inward and outward radial motion, while $\pm_\theta$ corresponds to polar angle oscillation directions. These expressions provide the foundation for subsequent calculations of innermost stable circular orbits and
accretion disk radiation profiles.

\subsection{ Key quantities of time-like geodesics}
Assuming that all massive constituent particles of the thin accretion disk move along equatorial circular timelike geodesics ($Q = 0$, $\theta = \pi/2$), we first derive the key physical quantities characterizing these orbits, which are essential for the subsequent analysis of the accretion disk observables.

The motion of a free test particle in curved spacetime is governed by the geodesic equation, which takes the form:
\begin{align}
\frac{d^2 x^\rho}{d\tau^2} + \Gamma^\rho_{\alpha\beta} \frac{dx^\alpha}{d\tau} \frac{dx^\beta}{d\tau} = 0.\label{eq:18}
\end{align}

To derive the angular velocity of circular motion, we impose the following dynamical conditions: the particle is confined to the equatorial plane, with vanishing radial and polar velocity and acceleration components ($\dot{r} = 0$, $\ddot{r} = 0$, $\dot{\theta} = 0$).

Taking $\rho = r$ (the radial direction) and substituting $\ddot{r} = 0$ into the geodesic Eq.~\eqref{eq:18}, we obtain the radial component of the geodesic equation in terms of Christoffel symbols:
\begin{align}
   \Gamma^r_{\alpha\beta} u^\alpha u^\beta = 0 ,  
\end{align}
where $u^\alpha = dx^\alpha/d\tau$. Expanding the non-vanishing terms (only the $t$ and $\phi$ directions have velocity):
\begin{align}
    \Gamma^r_{tt} (u^t)^2 + 2\Gamma^r_{t\phi} u^t u^\phi + \Gamma^r_{\phi\phi} (u^\phi)^2 = 0 . 
      \label{eq:21}
\end{align}

This equation gives us a quadratic equation for the angular velocity $\Omega \equiv u^\phi/u^t$. Solving this equation yields the analytical expression for the angular velocity of a test particle in equatorial circular motion:

\begin{align}
    \Omega = \frac{-\partial_r g_{t\phi} \pm \sqrt{\left(\partial_r g_{t\phi}\right)^2 - \left(\partial_r g_{tt}\right)\left(\partial_r g_{\phi\phi}\right)}}{\partial_r g_{\phi\phi}}. 
    \label{eq:22}
\end{align}

The plus sign corresponds to prograde equatorial orbits co-rotating with the black hole spin, while the minus sign corresponds to retrograde orbits counter-rotating relative to the black hole spin.

Substituting the expression for $\Omega$ into Eq.~\eqref{eq:21} and using $\Omega \equiv u^\phi / u^t$ restructures the equation into a form with $u^t$ and $\Omega$ as variables. Imposing the equatorial circular motion constraints, this yields the explicit closed-form expression for the time component $u^t$ of the four-velocity:

\begin{align}
    u^t  = \dfrac{1}{\sqrt{-\left(g_{tt} + 2g_{t\phi}\Omega + g_{\phi\phi}\Omega^2\right)}} 
    \label{eq:23} 
\end{align}

Combining Eq.~\eqref{eq:6} and Eq.~\eqref{eq:7}, we substitute $u^\phi$ and $u^t$ for $p^\phi$ and $p^t$ respectively, with the relation $u^\phi = \Omega u^t$. Then we obtain explicit closed-form expressions for $E$ and $L$,
\begin{align}
E &= -\frac{g_{tt} + g_{t\phi}\Omega}{\sqrt{-g_{tt} - 2g_{t\phi}\Omega - g_{\phi\phi}\Omega^2}},  
 \label{eq:24} \\
L &= \frac{g_{t\phi} + g_{\phi\phi}\Omega}{\sqrt{-g_{tt} - 2g_{t\phi}\Omega - g_{\phi\phi}\Omega^2}}.
\label{eq:25} 
\end{align}

We now determine the innermost stable circular orbit (ISCO) radius $r_{\text{ISCO}}$, which defines the inner boundary of the accretion disk. As derived in Eq.~\eqref{eq:14}, radial orbital motion is fully governed by the radial effective potential $\mathcal{R}(r)$. This critical radius is defined by the orbital stability condition that the second radial derivative of the radial effective potential $\mathcal{R}(r)$ vanishes:

\begin{align}
    \left. \frac{d^2 \mathcal{R}(r)}{dr^2} \right|_{r_{\text{ISCO}}} = 0.
\end{align}
Unlike the vacuum Kerr case, the modified spacetime does not admit a closed-form solution for $r_{\text{ISCO}}$, necessitating a numerical root-finding procedure \cite{Chen2025}.

\begin{figure}[htbp]
    \centering
\includegraphics[scale=0.5]{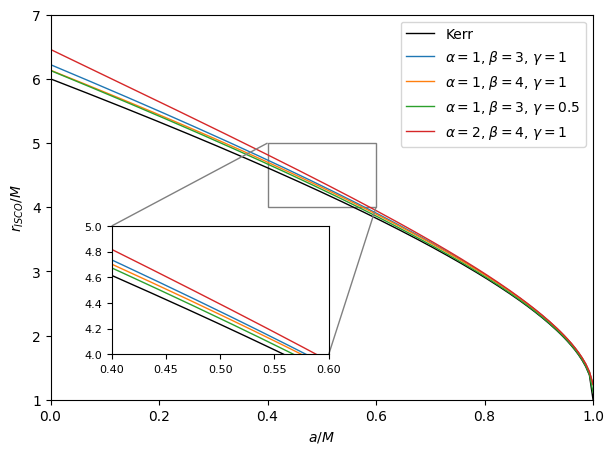}
\caption{ISCO radius versus spin $a/M$ and dark matter halo parameters $(\alpha, \beta, \gamma)$ by setting $\rho_s M^2 = 10^{-4},\; r_s = 30M$. Dark matter halos shift $r_{\text{ISCO}}$ outward relative to vacuum Kerr; the inset shows subtle model differences for $a/M \in [0.4, 0.6]$.}
\label{phib}
\end{figure}

Fig.~\ref{phib} presents the ISCO radius $r_{\rm ISCO}$ as a function of the dimensionless spin parameter $a/M$ for various dark matter halo parameter sets $(\alpha, \beta, \gamma)$. All curves exhibit a monotonically decreasing trend: as $a/M$ increases, $r_{\rm ISCO}/M$ decreases and approaches unity in the extremal limit ($a/M \to 1$), consistent with the vacuum Kerr result. This behavior reflects the well-known ``compactification'' effect of rapidly rotating black holes on stable circular orbits, and the inclusion of a dark matter halo does not modify this underlying physical mechanism.

Furthermore, all curves with dark matter halos lie above the standard vacuum Kerr curve, indicating that the presence of dark matter shifts the ISCO radius to larger values, which become more prominent for black holes with lower spins. 

\subsection{Key quantities of null geodesics}

Photons, as massless test particles ($\mu = 0$), follow null geodesics. Although the thin accretion disk model is constructed from timelike geodesics, the ray-tracing techniques employed for black hole imaging critically depend on precise photon trajectory calculations. It is therefore necessary to introduce the characteristic radii governing photon dynamics in the ray-tracing framework. 

In this rotating black holes with dark matter halos Eq.~(\ref{metric}), the zeros of the radial potential $\mathcal{R}(r)$ classify photon orbits into two topologically distinct families: photons are either captured by the black hole event horizon or escape to infinity. These critical roots govern the asymptotic behavior of photon trajectories, thereby imposing fundamental constraints on the observational signatures of the black hole shadow and photon ring. These critical roots are obtained by solving the equation $\mathcal{R}(r) = 0$, whose explicit form is given by:

\begin{align}
    r^4 + \mathcal{A} r^2 + \mathcal{B} r + \mathcal{C} = 0, \label{cr}
\end{align}
where

\begin{equation}
\begin{aligned}
A &= a^2 -\eta - \lambda^2 , \\
B &= 2m(r)\left[\eta + (\lambda - a)^2\right], \\
C &= -a^2 \eta .
\end{aligned}
\label{eq:metric_part2}
\end{equation}

The radial equation is recast in terms of coefficients $A$, $B$, and $C$ to facilitate comparison with established Kerr black hole results. However, this equation deviates from the standard quartic polynomial structure, as the coefficient \(B\) contains the mass function \(m(r)\), which implicitly depends on the radial coordinate \(r\) via the dark matter density profile. Specifically, this nonlinear coupling makes a closed-form analytical solution impossible. Therefore, numerical methods are essential. 

We now focus on determining the radii of the photon shell boundary in the equatorial plane \(\tilde{r}_{\pm}\) of this rotating black holes with dark matter halos. The photon shell constitutes a trapping region where photons are transiently trapped by spacetime curvature, and it hosts unstable bound orbits: photons neither plunge irreversibly into the event horizon nor escape to infinity, until infinitesimal perturbations eventually drive them toward either capture or scattering. The radii \(\tilde{r}\) of these bound photon orbits are obtained by solving:
\begin{align}
    \left. \mathcal{R}(r) \right|_{\tilde{r}} = \left. \frac{d\mathcal{R}(r)}{dr} \right|_{\tilde{r}} = 0 .
\end{align}

Substituting $\mathcal{R}(r)$ into above equations, we get

\begin{align}
\left(\tilde{r}^2 + a^2 - a\lambda\right)^2 &= \Delta(\tilde{r}) \left[\mu^2 \tilde{r}^2 + \eta + (\lambda - a)^2\right],  
    \label{eq:30} \\
4\tilde{r}\left(\tilde{r}^2 + a^2 - a\lambda\right) &= \Delta'(\tilde{r}) \left[\mu^2 \tilde{r}^2 + \eta + (\lambda - a)^2\right] \nonumber\\& \quad+ 2\mu^2 \tilde{r}\,\Delta(\tilde{r}) . 
    \label{eq:31} 
\end{align}

For photons ($\mu = 0$), solving the system of Eq.~(\ref{eq:30}) and (\ref{eq:31}) yields $\tilde{\lambda}$,

\begin{align}
    \tilde{\lambda}_{\mu=0} = \frac{1}{a}\left[\tilde{r}^2 + a^2 - \frac{4\tilde{r}\,\Delta(\tilde{r})}{\Delta'(\tilde{r})}\right] ,
    \label{eq:32} 
\end{align}
where
\begin{align}
    \Delta'(\widetilde{r}) = 2\left[ \widetilde{r} - m(\widetilde{r}) - \widetilde{r} m'(\widetilde{r}) \right]. 
\end{align}
By substituting Eq.~\eqref{eq:32} into Eq.~\eqref{eq:30} and \eqref{eq:31}, we have

\begin{align}
     \widetilde{\eta} = \frac{16\tilde{r}^2 \Delta(\tilde{r})}{\left[\Delta'(\tilde{r})\right]^2} - \frac{1}{a^2} \left[ \tilde{r}^2 - \frac{4\tilde{r}\Delta(\tilde{r})}{\Delta'(\tilde{r}) \vphantom{\left[\Delta'(\tilde{r})\right]^2}} \right]^2 . 
     \label{eq:35} 
\end{align}

For equatorial photon orbits with a vanishing Carter constant \(Q = 0\) (corresponding to \(\tilde{\eta} = 0\)), substituting this condition into Eq.~\eqref{eq:35} yields two radii \(\tilde{r}_{\pm}\). Both radii lie strictly outside the event horizon \(r_+\). Of these two roots, \(\tilde{r}_{-}\) and \(\tilde{r}_{+}\) mark the inner and outer boundaries of the shell, confining all bound photon orbits to the radial interval \([\tilde{r}_{-}, \tilde{r}_{+}]\), which is a thin spherical layer enveloping the rotating black holes with dark matter halos. 

\begin{figure}
    \centering
    \includegraphics[scale=0.55]{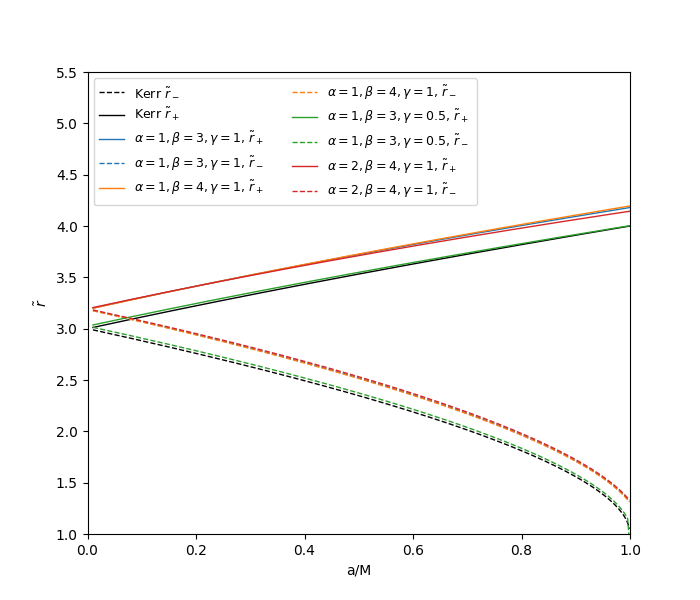}
    \caption{The inner (dashed curves, ${\tilde r}_-$) and outer (solid curves, ${\tilde r}_+$) boundary of photon shell as a function of black hole spin $a/M$ for rotating dark matter black hole, with different combinations of halo parameters $\alpha,\beta,\gamma$=(1,3,1), (1,4,1), (1,3,0.5), (2,4,1), while we set $\rho_sM^2=10^{-4}$ and $r_s/M=30$. The Kerr case is also shown as a reference. }
\label{phs}
\end{figure}

Fig.~\ref{phs} illustrates the inner $\tilde r_-$ and outer $\tilde r_+$ boundaries of the photon shell as functions of the black hole spin parameter for several representative dark matter halo profiles,
$(\alpha,\beta,\gamma)=(1,3,1)$, $(1,4,1)$, $(1,3,0.5)$, and $(2,4,1)$, while the halo density and scale radius are fixed to
$\rho_s M^2=10^{-4}$ and $r_s/M=30$, respectively.
For comparison, the corresponding Kerr results are also shown. From Fig.~\ref{phs}, we can observe that both the inner and outer boundaries of the photon shell are shifted outward relative to the Kerr case. This behavior reflects the additional gravitational field generated by the dark matter halo. Since the effective mass function $m(r)$ increases with radius, the spacetime curvature experienced by photons becomes stronger than that of an isolated Kerr black hole. As a result, unstable spherical photon orbits can exist at larger radii, leading to an expansion of the entire photon shell.

Another important observation is that the parameter $\gamma$ produces the most substantial modification of the photon shell. The reason can be traced to its role in controlling the inner density cusp of the dark matter halo. Since the photon shell is located only a few gravitational radii outside the horizon, its properties are mainly determined by the dark matter distribution in the innermost region of the halo. A larger value of $\gamma$ corresponds to a steeper central density profile and therefore a stronger gravitational contribution near the photon shell. In contrast, the parameters $\alpha$ and $\beta$ mainly govern the transition and outer regions of the halo profile and thus generate comparatively weaker effects on the photon shell structure.

An infinitesimal perturbation, whether from external gravitational fields or quantum effects, disrupts the unstable equilibrium, driving the photon toward either capture by the horizon or escape to infinity. This instability imprints a sharply defined critical curve on the observer's image plane or the black hole shadow, whose geometry and position depend sensitively on the spacetime parameters. We will show the critical curves with different black hole and halo parameters in Fig.~\ref{app1} of Sec.~\ref{sec5}.

\section{Emission properties of thin accretion disks}\label{sec4}

Based on the standard thin accretion disk model \cite{Novikov1973,Page1974,Shakura1973}, the local radiative flux $\mathcal{F}(r)$ quantifies the energy emitted per unit area per unit time at radial distance $r$ from the central black hole on the disk surface. Its
analytical expression is given by:

\begin{align}
    \mathcal{F}(r) = -\frac{\dot{m}}{4\pi\sqrt{-g}} \frac{\Omega_{,r}}{(E - \Omega L)^2} \int_{r_{ISCO}}^{r} (E - \Omega L) L_{,\tilde{r}} d\tilde{r}, 
    \label{eq:36}
\end{align}
where $\dot{m}$ denotes the mass accretion rate of the accretion disk.  The integration is performed from the ISCO radius $r_{\rm ISCO}$ to $r$, implying that $\mathcal{F}(r)$ is set to zero for $r<r_{\rm ISCO}$ within our model. Inside the ISCO, the accreting matter enters a plunging region and can no longer be described as a sequence of stable circular geodesics. Since our analysis is based on the standard thin accretion disk approximation, we take the ISCO as the inner edge of the disk and do not explicitly model the plunging region between the ISCO and the event horizon. All calculations in this work adopt a sub-Eddington accretion regime, where the accretion rate is well below the Eddington critical limit. Such an accretion condition satisfies all fundamental assumptions of the thin disk model, including geometrically thin disk configuration, local thermal radiation and viscous dissipation dominated energy release. The variable $g$ refers to the metric determinant of the three-dimensional subspace $(t, r, \phi)$ that characterizes the geometric properties of curved spacetime, with its specific form given by:

\begin{align}
    g = g_{rr}\!\left(g_{tt}g_{\phi\phi} - g_{t\phi}^2\right) .
\end{align}

Three critical orbital parameters, namely the angular velocity $\Omega$, specific energy $E$ and specific angular momentum $L$ of test particles moving around the black hole, are defined in Eq.~\eqref{eq:22}, Eq.~\eqref{eq:24} and Eq.~\eqref{eq:25}, respectively. The notations $\Omega_{,r}$ and $L_{,r}$ represent the first-order radial derivatives of orbital angular velocity and specific angular momentum, describing the radial variation gradients of these two orbital quantities across the accretion disk.
Consistent with the timelike geodesic assumptions in Sec.~\ref{sec2}, all accretion flows in this study are confined to the black hole equatorial plane ($\theta = \pi/2$). This constraint eliminates polar-angle-related cross terms in the metric, simplifying the calculation of $\Omega$, $E$, and $L$ and reducing the computational complexity
of the disk radiative flux in curved spacetime.

\begin{figure}[htbp]
    \centering
\includegraphics[scale=0.4]{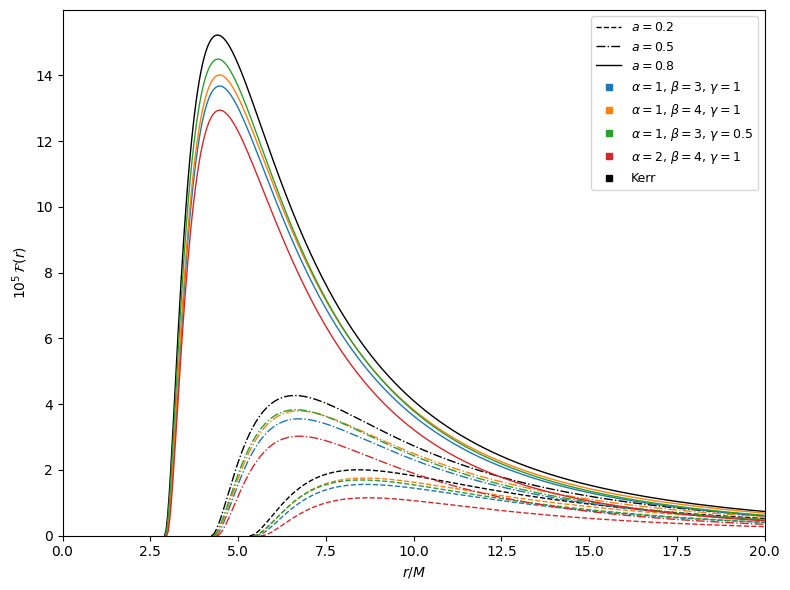}
\caption{The radiative flux \(10^5 \mathcal{F}(r)\) of a thin accretion disk as a function of \(r/M\) for different black hole spins \(a\) and dark matter halo parameters \((\alpha, \beta, \gamma)\) by setting $\rho_s M^2 = 10^{-4},\; r_s = 30M$. The Kerr black hole result is also shown for reference.}
\label{flux}
\end{figure}

Fig.~\ref{flux} presents the radial profiles of normalized radiative flux under different parameter combinations, including various black hole spin parameters $a$ and three dark matter halo parameters $\alpha$, $\beta$ and $\gamma$. To improve the visibility and comparability of the results, all flux values are scaled by a factor of $10^5$ (for a unit accretion rate), as the raw radiative flux values are extremely small and would otherwise obscure the relative trends between different parameter sets.

It also illustrates how black hole spin $a$ and dark matter halo parameters $(\alpha, \beta, \gamma)$ collectively shape the
radiative flux $10^5 \mathcal{F}(r)$ of a thin accretion disk. As the spin $a$ increases, the peak flux rises
sharply while its position shifts inward, a direct signature of the stronger frame-dragging effect in rapidly rotating black holes that accelerates inner-disk matter to higher energies and enhances radiative efficiency. Dark matter parameters suppress the peak flux. Notably, all curves converge at \(r/M \gtrsim 15\), indicating that in the far outer disk, radiation is dominated by the global accretion rate, with the local influences of black hole spin and dark matter distribution becoming negligible.

Assuming local thermal equilibrium, the disk temperature $T(r)$ is related to the radiative flux via the Stefan-Boltzmann law:

\begin{align}
    T = \sqrt[4]{\mathcal{F}(r)/\sigma}, 
\end{align}
where $\sigma$ is the Stefan-Boltzmann constant.

\begin{figure}[htbp]
    \centering
\includegraphics[scale=0.4]{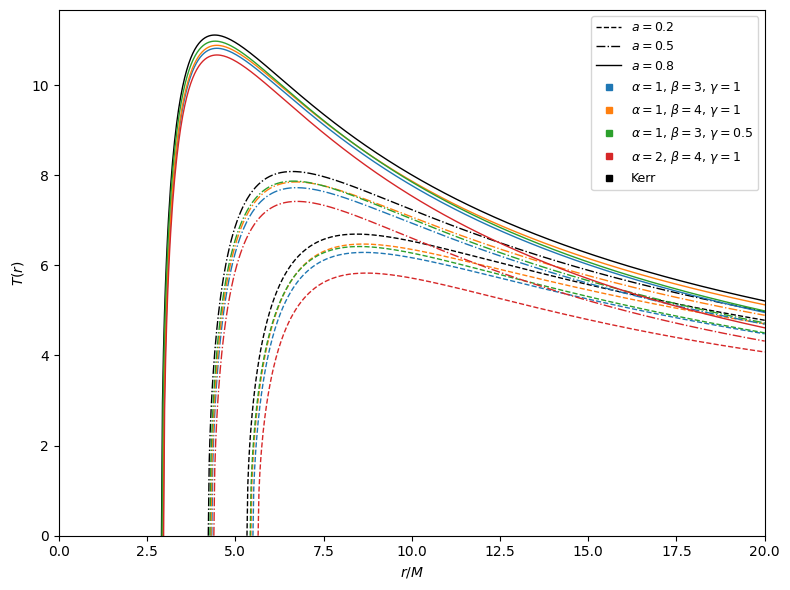}
\caption{Temperature distribution $T(r)$ of a thin accretion disk versus $r/M$ for various combinations of black hole spin $a$ and dark matter parameters $(\alpha,\beta,\gamma)$, by setting $\rho_s M^2 = 1\mathrm{e}{-4}$, $r_s = 30M$. All plotted values are multiplied by $10^2$, and results for the Kerr metric are included for comparison. }
\label{phib3}
\end{figure}

Fig.~\ref{phib3} illustrates how black hole spin $a$ and dark matter halo parameters $(\alpha, \beta, \gamma)$ shape the disk
temperature distribution $T(r)$. As spin $a$ increases from $0.2$ to $0.8$, the peak temperature rises sharply and
shifts inward, mirroring the trend observed in the radiative flux. This reflects stronger frame-dragging that
intensifies inner-disk frictional heating. Dark matter parameters suppress the peak temperature and shift it
outward, consistent with their effect on the radiative flux. All curves converge at $r/M \gtrsim 15$, meaning outer-disk temperature is dominated by the global accretion rate, with local spin and dark matter effects
negligible at large radii.

Another physically relevant quantity we study is the differential luminosity of the thin accretion disk, which is observable. It takes the form:

\begin{align}
    \frac{d\mathcal{L}_\infty}{d\ln r} = 4\pi r \sqrt{-g} E \mathcal{F}(r).
\end{align}

\begin{figure}[htbp]
    \centering
\includegraphics[scale=0.4]{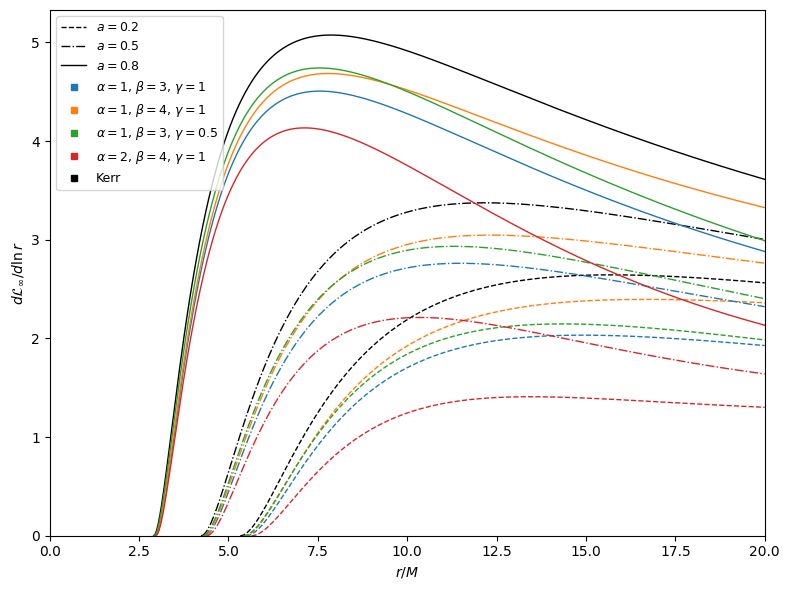}
\caption{Radial distribution of \(d\mathcal{L}_\infty/d\ln r\) for a thin accretion disk as a function of \(r/M\) under various combinations of black hole spin \(a\) and dark matter halo parameters \((\alpha,\beta,\gamma)\) by setting $\rho_s M^2 = 10^{-4}$, $r_s = 30M$. All plotted values are multiplied by  $10^2$, and results for the Kerr metric are included for comparison.}
\label{phib4}
\end{figure}

Fig.~\ref{phib4} shows the differential luminosity $d\mathcal{L}_\infty/d\ln r$, which follows analogous trends to the flux and temperature profiles: higher spin amplifies the luminosity peak and shifts it inward, while dark matter parameters suppress the peak flux, and their effects become more prominent in outer regions.

Furthermore, we investigated the spectral luminosity distribution characteristics of the thin accretion disk. The spectral luminosity
observed at infinity $\mathcal{L}_{\nu,\infty}$ is given by \cite{l1,l2,l3,l4,Li2025}

\begin{align}
    \nu \mathcal{L}_{\nu,\infty} = \frac{60}{\pi^3} \int_{r_{\rm isco}}^{\infty} \frac{\sqrt{g}\,E\,\left(u^t y\right)^4}{\exp\left[u^t y / \mathcal{F}^{1/4}\right] - 1} dr, 
\end{align}
where $\nu$ is the frequency, $y = h\nu/kT$, $h$ is the Planck constant, $k$ is the Boltzmann constant.

\begin{figure}[htbp]
    \centering
\includegraphics[scale=0.4]{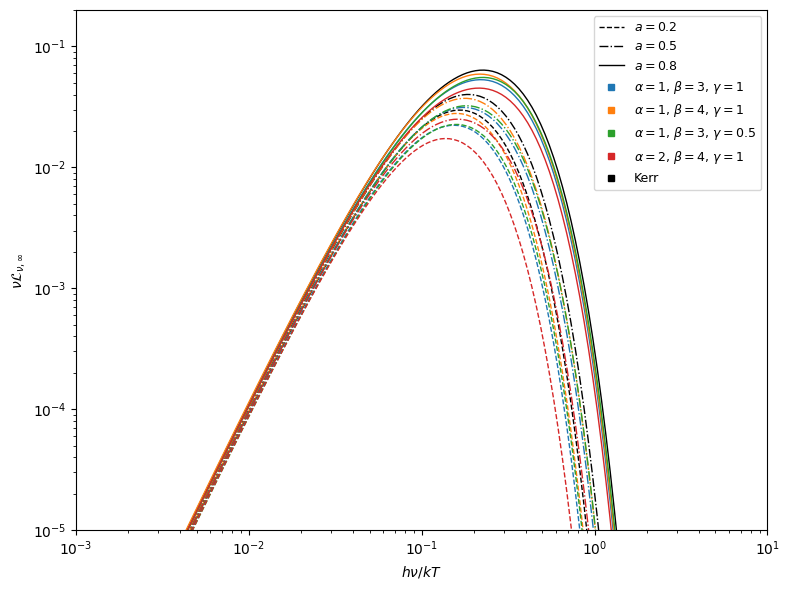}
\caption{Variation of $\nu \mathcal{L}_{\nu,\infty}$ with $h\nu/kT$ for a thin accretion disk under different combinations of black hole spin $a$ and  dark matter halo parameters  $(\alpha,\beta,\gamma)$, by setting $\rho_s M^2 = 10^{-4},\ r_s = 30M$. The presence of dark matter suppresses high-energy radiation and reduces the spectral peak. }
\label{phib5}
\end{figure}

Fig.~\ref{phib5} shows the variation of the spectral luminosity distribution \(\nu\mathcal{L}_{\nu,\infty}\) with the dimensionless frequency \(h\nu/kT\) under different combinations of black hole spin \(a\) and dark matter halo parameters \((\alpha, \beta, \gamma)\).
Fig.~\ref{phib5} demonstrates that the dark matter halo notably suppresses high-energy radiative emission and reduces the amplitude of the spectral peak. All spectra converge at low dimensionless frequencies, revealing that outer-disk radiative characteristics are insensitive to dark matter halo parameters.

\section{Bolometric images of thin accretion disks}\label{sec5}

\subsection{Ray-tracing method}

To generate black hole images, we employ the backward ray-tracing formalism. Within this framework, null geodesics are traced backward from the observer's viewing screen toward the black hole. The emergent intensity at each screen pixel is computed from the intersection of the ray with the thin accretion disk. The photon conserved quantities $(\lambda, \eta)$ uniquely determine its apparent sky position $(\alpha, \beta)$ \cite{Cunningham1973}, which are defined for a static distant observer placed at radial coordinate $r_o$ relative to black hole, with fixed polar inclination angle $\theta_o$. They are given by \cite{Cunningham1973}

\begin{align}
    \alpha = \lim_{r_0 \to \infty} \left( - r_0^2 \sin\theta_0 \frac{d\phi}{dr} \right), \quad
\beta = \lim_{r_0 \to \infty} \left( r_0^2 \frac{d\theta}{dr} \right). 
\end{align}

For a distant observer at $r_o \to \infty$, the spacetime metric admits asymptotic simplification. The corresponding approximate metric forms read $\Delta = r^2 + a^2 - 2Mr \approx r^2$ and $\Sigma = r^2 + a^2\cos^2\theta \approx r^2$. Substituting the radial derivative relations $d\theta/dr = p^\theta/p^r$ and $d\phi/dr = p^\phi/p^r$, together with Eq.~\eqref{eq:9}, \eqref{eq:10}, \eqref{eq:16} and \eqref{eq:17}, yields
\begin{align}
    \alpha_0 = -\frac{\lambda}{\sin\theta_0}, \quad
\beta = \pm_o \sqrt{\eta + a^2 \cos^2\theta_0 - \lambda^2 \cot^2\theta_0} ,
\end{align}

where $\pm_o$ is the sign of $\cos\theta_0$, reversely

\begin{align}
    \lambda = -\alpha \sin\theta_0, \quad
\eta = (\alpha^2 - a^2) \cos^2\theta_0 + \beta^2. 
\label{eq:42}
\end{align}

The coordinates $\alpha$ and $\beta$ encode the apparent sky position of each photon for distant observers, with their values fully parameterized by two conserved orbital integrals $\lambda$ and $\eta$. While $\alpha$ records the orbital rotational projection of photon orbital motion, $\beta$ encapsulates the combined contributions of polar orbital motion and black hole spin. These sky coordinates provide the mapping between photon trajectories and the black hole shadow.

Once the unique mapping relation between the viewing screen coordinates $(\alpha, \beta)$ and photon conserved integrals $(\lambda, \eta)$ is established through Eq.~\eqref{eq:42}, integrating the geodesic equations yields the complete null ray path from the observer to the emission region. For rotating black holes with dark matter halos, the geodesic equations are recast as definite integrals amenable to numerical evaluation.

Multiplying both sides of Eqs.~\eqref{eq:16} and \eqref{eq:17} by the factor $E/\Sigma$, we obtain

\begin{align}
p^{r} &= \frac{E}{\Sigma} \pm_{r} \sqrt{\mathcal{R}(r)} = \frac{dr}{d\lambda} , \label{eq:43}\\
p^{\theta} &= \frac{E}{\Sigma} \pm_{\theta} \sqrt{\Theta(\theta)} = \frac{d\theta}{d\lambda} ,  \label{eq:44}
\end{align}
where $\lambda$ is the affine parameter for null geodesic. By combining these two expressions and eliminating $\lambda$, we can construct the differential relation between $r$ and $\theta$:
\begin{equation}
\frac{dr}{d\theta} = \frac{dr}{d\lambda} \cdot \frac{d\lambda}{d\theta}
= \frac{\pm_{r} \sqrt{\mathcal{R}(r)}}{\pm_{\theta} \sqrt{\Theta(\theta)}}. 
\label{eq:dr_over_dtheta}
\end{equation}
Rearranging terms yields a unified differential form for integration:
\begin{equation}
\frac{dr}{\pm_{r}\sqrt{\mathcal{R}(r)}} = \frac{d\theta}{\pm_{\theta}\sqrt{\Theta(\theta)}}.
\label{eq:47}
\end{equation}

Based on the unified differential equality derived in Eq.~\eqref{eq:47}, two definite path integrals are constructed to quantify radial and polar propagation weights for null trajectories:

\begin{align}
I_r &= \int_{r_e}^{r_o} \frac{dr}{\pm_{r}\sqrt{\mathcal{R}(r)}} \label{eq:I_r}  ,\\
G_\theta &= \int_{\theta_e}^{\theta_o} \frac{d\theta}{\pm_{\theta}\sqrt{\Theta(\theta)}}.  
\end{align}

Here, $r_e$ denotes the radial coordinate of the photon emission source, and $r_o$ represents the radial coordinate of the distant observer. Similarly, $\theta_e$ is the polar angle at the emission point, while $\theta_o$ stands for the polar viewing angle of the observer. The integral $I_r$ quantifies the accumulated path weight of the photon’s radial propagation from the emission source to the observer, and $G_\theta$ characterizes the path weight associated with the polar angular oscillation of the photon. The core identity can be directly derived from the differential equivalence relation.

\begin{equation}
I_r = G_\theta . 
\label{eq:integral_equivalence}
\end{equation}

This identity exposes an intrinsic geometric equivalence between radial and polar orbital evolution. This mutual convertibility of radial and polar path lengths originates from the stationary axisymmetric spacetime, a geometric feature that imposes strict kinematic constraints on all null particle trajectories.

A photon travels from the emission azimuthal angle $\phi_\mathrm{e}$ to the receiving azimuthal angle $\phi_0$, and the total azimuthal deflection angle is defined as the difference of the two coordinate values.

\begin{equation}
\Delta \phi = \phi_0 - \phi_\mathrm{e} = \int_{\lambda_\mathrm{e}}^{\lambda_0} \frac{\mathrm{d}\phi}{\mathrm{d}\lambda} \mathrm{d}\lambda = \int_{\lambda_\mathrm{e}}^{\lambda_0} p^\phi \mathrm{d}\lambda. 
\label{eq:51}
\end{equation}

The azimuthal momentum $p^\phi$ can be derived from Eq.~\eqref{eq:9}. Using the identity $r^2 + a^2 = \Delta + 2 r m(r)$,
we obtain the identity
\begin{equation}
r^2 + a^2 - a \lambda = \Delta + 2 r m(r) - a \lambda.  
\label{eq:52}
\end{equation}

Substituting Eqs.~\eqref{eq:9} and \eqref{eq:52} to \eqref{eq:51} yields

\begin{equation}
\Delta \phi = \int_{\lambda_\mathrm{e}}^{\lambda_0} \frac{E}{\Sigma} \cdot \frac{a}{\Delta} \bigl(2 r m(r) - a \lambda\bigr) \mathrm{d}\lambda + \lambda\int_{\lambda_\mathrm{e}}^{\lambda_0} \frac{E}{\Sigma} \csc^2 \theta \, \mathrm{d}\lambda . 
\label{eq:azimuth_deflect_integral}
\end{equation}

Using Eqs.~\eqref{eq:43} and \eqref{eq:43} to eliminate $d\lambda$ yields

\begin{equation}
\Delta \phi = \int_{r_\mathrm{e}}^{r_0} \frac{a\bigl(2 r m(r)-a\lambda\bigr)}{\pm_r \Delta \sqrt{\mathcal{R}(r)}} \mathrm{d}r + \lambda\int_{\theta_\mathrm{e}}^{\theta_0} \frac{\csc^2\theta}{\pm_\theta \sqrt{\Theta(\theta)}} \mathrm{d}\theta .
\end{equation}

We define the radial and polar deflection integrals:

\begin{align}
I_\phi &= \int_{r_\mathrm{e}}^{r_0} \frac{a\bigl(2 r m(r) - a\lambda\bigr)}{\pm_r \Delta \sqrt{\mathcal{R}(r)}} \mathrm{d}r ,\label{eq:int_radial_phi}  \\
G_\phi &= \int_{\theta_\mathrm{e}}^{\theta_0} \frac{\csc^2 \theta}{\pm_\theta \sqrt{\Theta(\theta)}} \mathrm{d}\theta . 
\end{align}

$I_\phi$ is a definite integral evaluated solely with respect to the radial coordinate, which physically quantifies the total intrinsic orbital rotational deflection accumulated along the photon's radial trajectory.
This deflection originates from the spacetime frame-dragging effect generated by the central rotating black holes with dark matter halos. $G_\phi$ characterizes the additional azimuthal deviation, which originates from the photon’s off-equatorial motion and oscillating polar angle. Then, we have

\begin{equation}
\Delta \phi = \phi_0 - \phi_\mathrm{e} = I_\phi + \lambda G_\phi. 
\end{equation}

This expression decomposes the complicated integral over the affine parameter into two separate definite integrals with respect to the radial and polar coordinates, which greatly simplifies the numerical calculation of photon deflection in the rotating black holes with dark matter halos.

The total coordinate time delay of photon propagation is defined as the difference between the observer’s reception time $t_0$ and the source’s emission time $t_\mathrm{e}$.

\begin{equation}
\Delta t = t_0 - t_\mathrm{e} = \int_{\lambda_\mathrm{e}}^{\lambda_0} \frac{\mathrm{d}t}{\mathrm{d}\lambda} \,\mathrm{d}\lambda = \int_{\lambda_\mathrm{e}}^{\lambda_0} p^t \mathrm{d}\lambda. 
\label{eq:58}
\end{equation}

Algebraic rearrangement is carried out by adopting the geometric identity listed below.

\begin{equation}
r^2 + a^2 = \Delta + 2 r m(r). 
\label{eq:59}
\end{equation}

Combining Eqs.~\eqref{eq:10} and ~\eqref{eq:59} gives

\begin{equation}
p^t = \frac{E}{\Sigma} \left[ \frac{2 r m(r)\bigl(r^2 + a^2 - a \lambda\bigr)}{\Delta} + r^2 + a^2 \cos^2\theta \right]. 
\label{eq:p_t_canonical}
\end{equation}

Substitute the polynomial into Eq.~\eqref{eq:58}, expand and reorganize the integrand, separate purely radial terms and purely polar trigonometric terms, and recast the expression into a form with fully decoupled radial and polar variables.

\begin{align}
\Delta t =& \int_{\lambda_{\mathrm{e}}}^{\lambda_{0}} \frac{E}{\Sigma} \cdot \frac{r^{2} \Delta + 2 r m(r)\left(r^{2}+a^{2}-a \lambda\right)}{\Delta} \mathrm{d}\lambda \nonumber \\
&+ \int_{\lambda_{\mathrm{e}}}^{\lambda_{0}} \frac{E}{\Sigma} a^{2} \cos^{2}\theta \,\mathrm{d}\lambda. 
\label{eq:61}
\end{align}

Using again Eqs.~\eqref{eq:43} and \eqref{eq:43}, we have

\begin{align}
\Delta t =& \int_{r_{\mathrm{e}}}^{r_{0}} \frac{r^{2} \Delta + 2 r m(r)\left(r^{2}+a^{2}-a \lambda\right)}{\Delta \cdot \pm_{r} \sqrt{\mathcal{R}(r)}} \mathrm{d}r \nonumber \\
&+ a^{2} \int_{\theta_{\mathrm{e}}}^{\theta_{0}} \frac{\cos^{2}\theta}{\pm_{\theta} \sqrt{\Theta(\theta)}} \mathrm{d}\theta. 
\label{eq:62}
\end{align}

The radial gravitational propagation time delay integral and the polar spin-coupled time delay integral are defined, and their combination yields the decomposition expression of the total coordinate time delay.

\begin{align}
I_t &= \int_{r_\mathrm{e}}^{r_0} \frac{r^2 \Delta + 2 r m(r)\bigl(r^2 + a^2 - a \lambda\bigr)}{\pm_r \Delta \sqrt{\mathcal{R}(r)}} \mathrm{d}r ,\label{eq:I_t_radial_time} \\
G_t &= \int_{\theta_\mathrm{e}}^{\theta_0} \frac{\cos^2\theta}{\pm_\theta \sqrt{\Theta(\theta)}} \mathrm{d}\theta ,\label{eq:G_t_polar_time} \\
\Delta t &= t_0 - t_\mathrm{e} = I_t + a^2 G_t \label{eq:total_time_delay}
. 
\end{align}

The integral $I_t$ corresponds to the fundamental time delay induced by spacetime curvature and gravitational redshift as photons propagate radially through the gravitational field, which only depends on radial trajectories and dominates the observed gravitational time delay at large distances. The integral $G_t$ describes the additional time offset arising from spacetime coupling of the rotating black holes with dark matter halos during the photon’s oscillatory polar motion, where the coefficient $a^2$ modulates the magnitude of spin-coupled delay; $G_t \equiv 0$ and the coupled delay vanishes identically for equatorial photon orbits. This decomposition separates the curvature and spin contributions, facilitating independent constraints on spacetime parameters.

This framework integrates radial, polar, azimuthal, and temporal geodesic evolution into a computationally tractable set of integrals, enabling the analysis of particle orbits, precession, and horizon crossing.

We employ line integration along photon trajectories as the core computational method. The angular integrals $(G_\theta, G_\phi, G_t)$ retain the same closed-form structure as in the vacuum Kerr case \cite{Gralla:2020a,Gralla:2020b}, whereas the radial integrals $(I_r, I_\phi, I_t)$ acquire additional terms from the dark matter halo–modified mass function $m(r)$.

For the standard vacuum Kerr spacetime, all radial integration problems admit closed-form analytical expressions, and the corresponding integral transformations are bidirectionally invertible. For the rotating black holes with dark matter halos, the radial integrals must be evaluated numerically. Inverting the functional mapping encoded within the integral system permits exact recovery of the photon emission spatial coordinates, given prior knowledge of the two conserved orbital integrals $(\lambda, \eta)$ associated with each null light ray.

Once the spatial coordinate of every radiative emission source point is solved, the associated radiative intensity is evaluated via a tailored thin disk emission distribution model introduced in Sec.~\ref{sec3}. Under this disk model, the radiative flux at any emission location depends uniquely on the radial source coordinate $r_e$, see Eq.~\eqref{eq:36}. All photon radiative contributions are subsequently mapped onto discrete pixel coordinates $(\alpha, \beta)$ on the observer's viewing screen.

Every circular radiation ring with radius $r_e$ lying on the equatorial plane of this thin accretion disk will generate multiple separate projections on the imaging plane, as light rays originating from these photon rings experience strong gravitational lensing and may undergo multiple deflections around the rotating black holes with dark matter halos, their trajectories are determined by the critical roots of the radial potential obtained from Eq.~\eqref{cr}. We classify these separated imaging zones based on how many times each light ray crosses the disk’s equatorial plane during its travel from emission point to observer, and name each classified zone an $n$-rank gravitational lensing band. The case of $n=0$ corresponds to direct imaging signals, for which light rays never intersect the equatorial plane at any stage of their propagation path. Light rays that cross the equatorial plane exactly one single time will form imaging signals confined within the $n=1$ lensing band. Increasing the lensing order $n$ yields three consistent trends for each lensing band: the observed radiative flux diminishes, the spatial angular extent of the band contracts, and the band's central position shifts progressively toward the critical photon curve defining the black hole shadow boundary. To assemble a fully superposed complete black hole radiation image, we have to sum up all radiation contributions delivered by light rays from all classified lensing band.

Lensing bands of order $n\geq3$ overlap almost entirely with the critical curve and carry drastically suppressed radiative flux magnitude relative to the low-order lensing bands with $n=0$, $n=1$, and $n=2$. For this reason, we only incorporate contributions from strips with $n=0$, $n=1$ and $n=2$ in all our numerical calculations. Combined with the radiation flux distribution at source positions $\mathcal{F}(r)$, the overall observed radiation flux $\mathcal{F}_o$ or bolometric image recorded on the imaging plane obeys the summation formula listed below:
\begin{equation}
\mathcal{F}_o(\alpha,\beta) = \sum_{n=0}^{2} \chi^4\left(r_s^{(n)},\alpha,\beta\right) \mathcal{F}_s\left(r_s^{(n)}\right).
\end{equation}
In this formula, parameter $\chi$ stands for the redshift modification coefficient, which is defined by the following equation:
\begin{equation}
\chi = \frac{1}{u^t(1 - \lambda\Omega)}.
\end{equation}
The explicit mathematical expression of term $u^t$ is derived and displayed in the orbital motion Eq.~\eqref{eq:23} demonstrated earlier within this paper.

In summary, the full backward photon ray-tracing simulation adopts a self-consistent closed-loop computational pipeline. Every discrete computational step within this framework relies on the analytical geodesic and integral relations derived in preceding sections.

\subsection{Bolometric images of rotating black holes with dark matter halos}

We optimized the classical $\textbf{aart}$\footnote{\url{https://github.com/iAART/aart}} \cite{Cardenas2023} code via the ray-tracing method, enabling it to numerically compute observable quantities and extending its applicability to non-Kerr black holes. We focused on analyzing the gravitational lensing bands and bolometric images of rotating black holes with dark matter halos as functions of black hole spin and inclination angle, and dark matter halo parameters.  Compared the imaging results with those of Kerr black holes, it will reveal the core differences induced by the presence of dark matter halo.

\begin{figure*}[htbp]
  \centering
    \includegraphics[scale = 0.5]{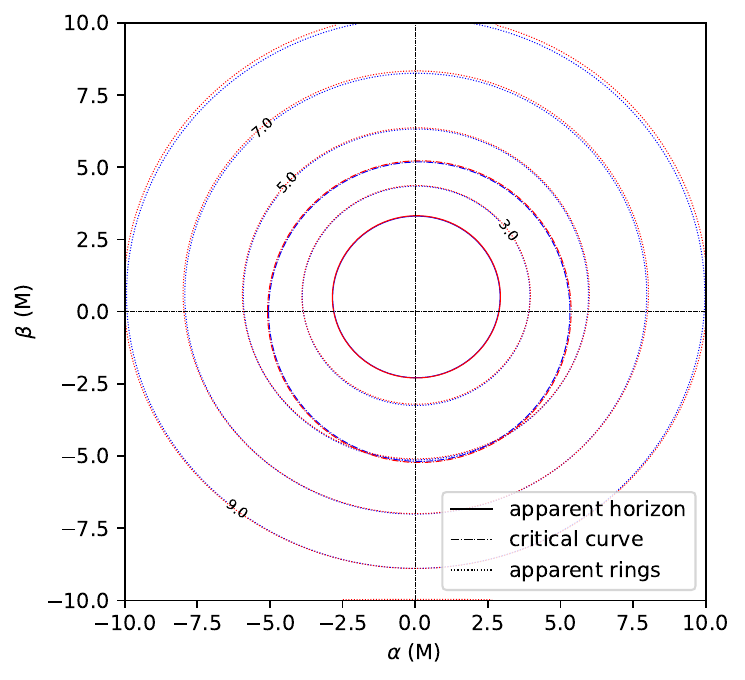}
    \includegraphics[scale = 0.5]{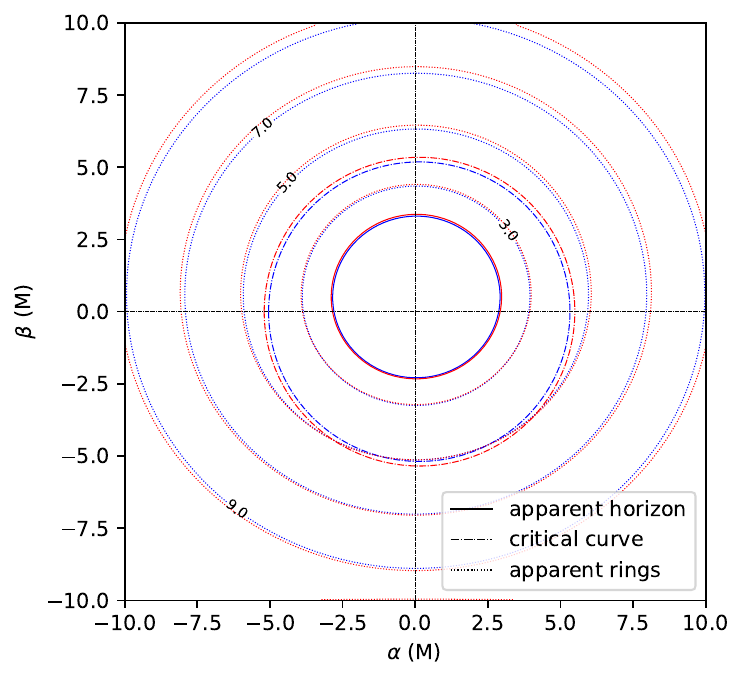}
    \includegraphics[scale = 0.5]{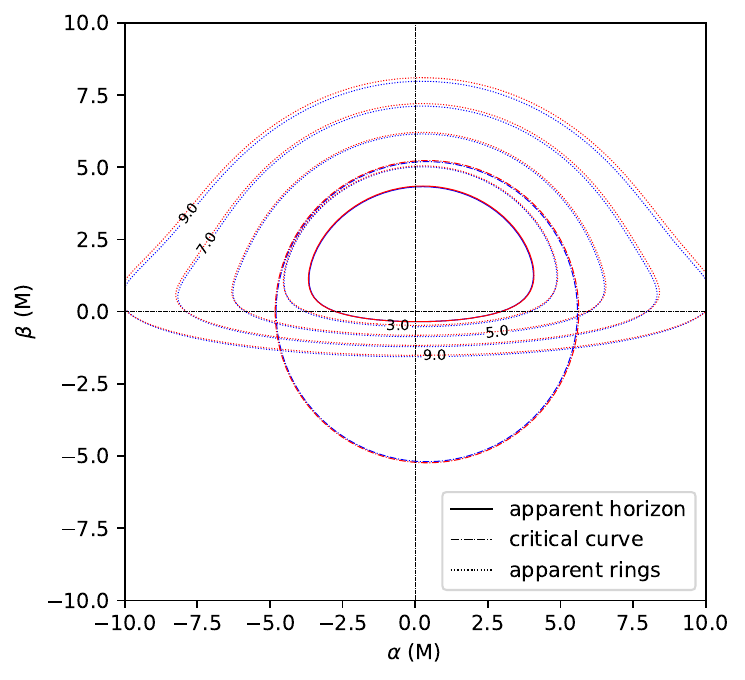}
    \includegraphics[scale = 0.5]{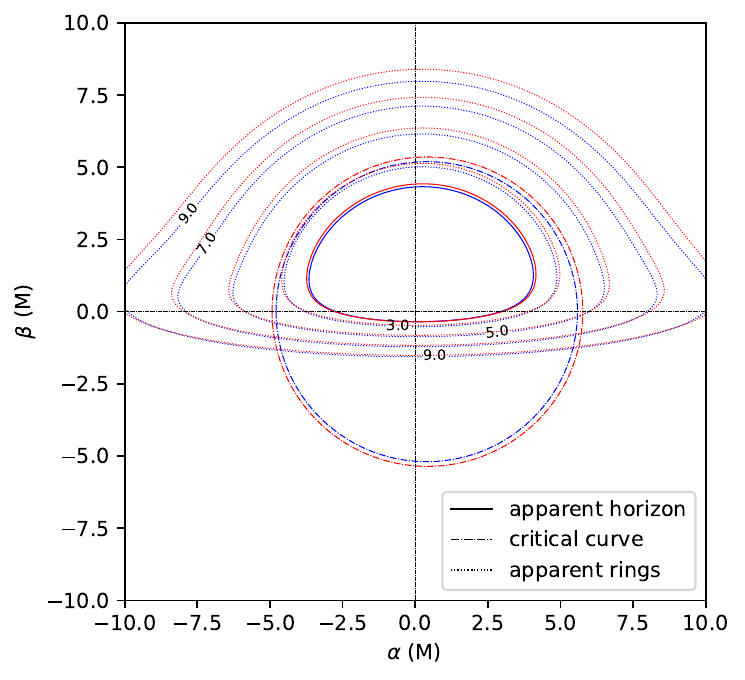}
    \includegraphics[scale = 0.5]{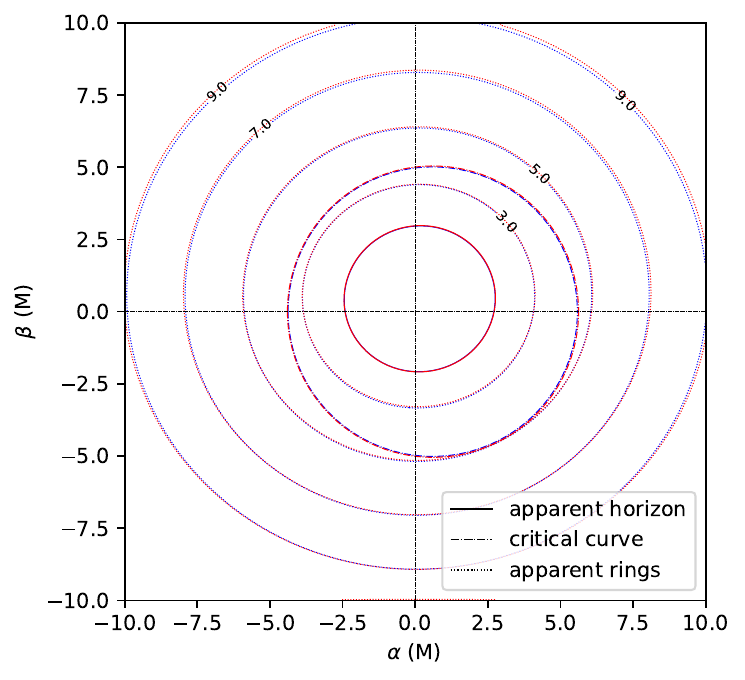}
    \includegraphics[scale = 0.5]{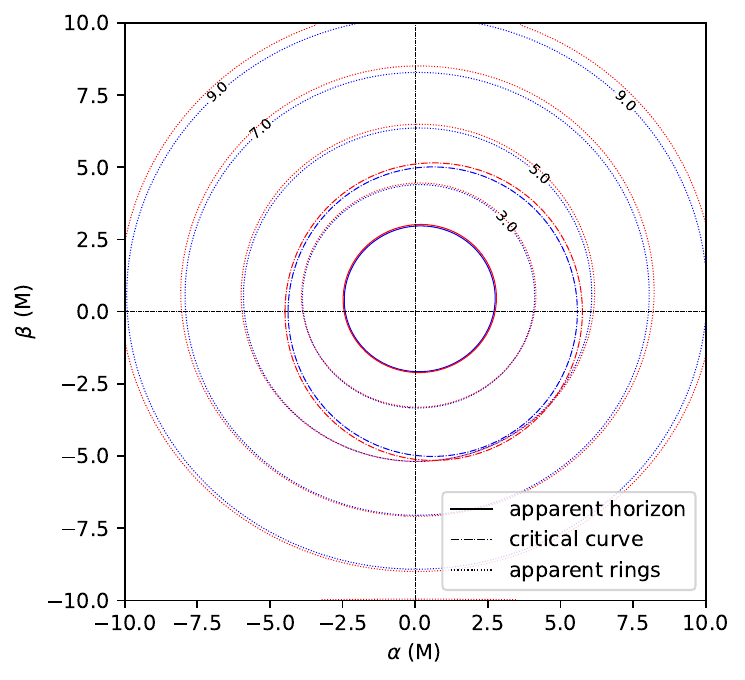}
    \includegraphics[scale = 0.5]{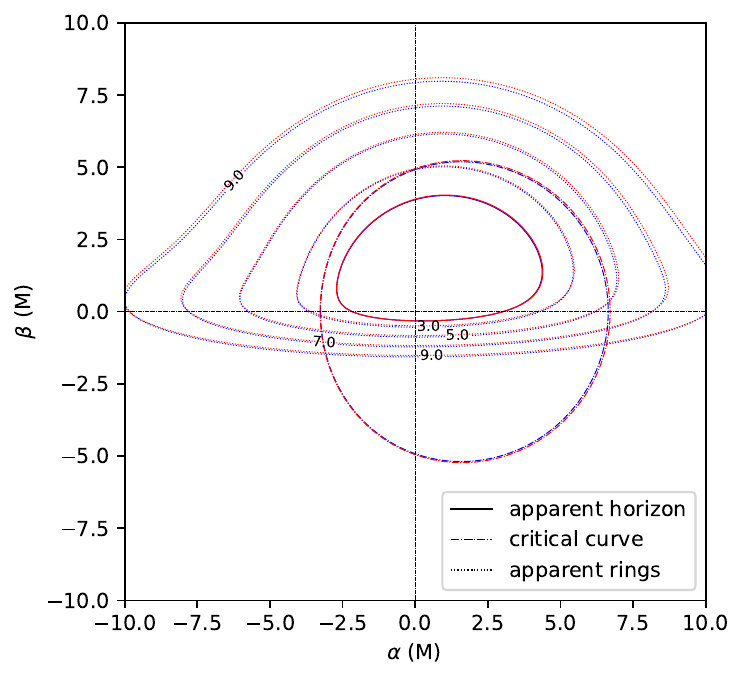}
    \includegraphics[scale = 0.5]{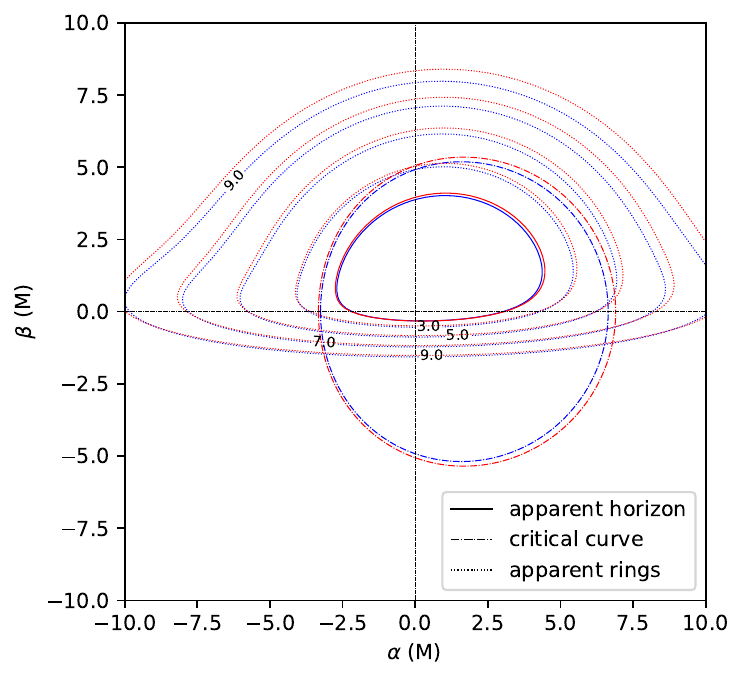}
\caption{Comparison of characteristic curves for Kerr black holes (blue curves) and rotating black holes with dark matter halos (red curves). The dark matter halo parameter set is $\alpha, \beta, \gamma =(1,3,0.5)$ for the left panel and $(1,3,1)$ for the right panel, while fixing $\rho_s M^2 = 2\times 10^{-5}$, $r_s = 30M$. All panels display the apparent horizon, $n=0$ apparent source rings (with radial coordinates $r_e/M=3,5,7,9$), and critical curves. The first to fourth rows correspond to four combinations of black hole spin and viewing inclination angle: $(a=0.2,\,i=20^\circ)$, $(a=0.2,\,i=80^\circ)$, $(a=0.8,\,i=20^\circ)$, and $(a=0.8,\,i=80^\circ)$, respectively. }
\label{app1}
\end{figure*}

In Fig.~\ref{app1},  we show the characteristic curves: apparent horizon, critical curve, and $n=0$ apparent rings, for Kerr black holes (blue curves) and rotating black holes with dark matter halos (red curves). We fix the dark matter halo parameter $\rho_sM^2=2\times10^{-5}$ and $r_s/M=30$ for all panels. The left column adopts the dark matter parameter set $\alpha,\beta,\gamma= (1,3,0.5)$ and the right column uses $\alpha,\beta,\gamma = (1,3,1)$. We select these two parameter combinations $(1, 3, 0.5)$ and $(1, 3, 1)$ based on the discussions on Fig.~\ref{phs}, which indicate that the parameter $\gamma$ exerts the most prominent influence on the black hole shadow. The four rows sequentially present four combinations of black hole spin and viewing inclination angle: $(a=0.2,i=20^\circ)$, $(a=0.2,i=80^\circ)$, $(a=0.8,i=20^\circ)$, and $(a=0.8,i=80^\circ)$.

Comparing the lensing band profiles of all eight sub-figures in Fig.~\ref{app1}, we can clearly distinguish the observational differences between the two types of black holes. Layered analysis of geometric structures shows that dark matter barely changes the projected size of the apparent horizon, and the red and blue solid lines for the apparent horizon almost overlap completely. By contrast, critical curves and multi-layer emission rings exhibit prominent outward expansion. For the same radial label, red contours always lie outside blue ones, and the expansion magnitude increases with the radial coordinate of the emission source and the dark matter halo parameter $\gamma$. The separation between red and blue contours is much larger for far-field source rings, which indicates that the spacetime correction induced by dark matter halo accumulates gradually with the orbital radius. Comparison between left and right columns reveals that all bands expand more significantly in the right panels with larger dark matter halo parameters $\gamma$.

Black hole spin and viewing inclination further modulate the imaging deformation caused by dark matter. Under low spin and inclination, the imaging structures of both black hole models are close to perfect circles, the spacing between red and blue bands distributes uniformly, and morphological discrepancies remain relatively weak. At high spin and inclination, images compressed along the negative $\beta$ direction into a teardrop shape with greatly enhanced asymmetry, and the outward expansion effect is further amplified.

\begin{figure*}[htbp]
  \centering
    \includegraphics[scale = 0.55]{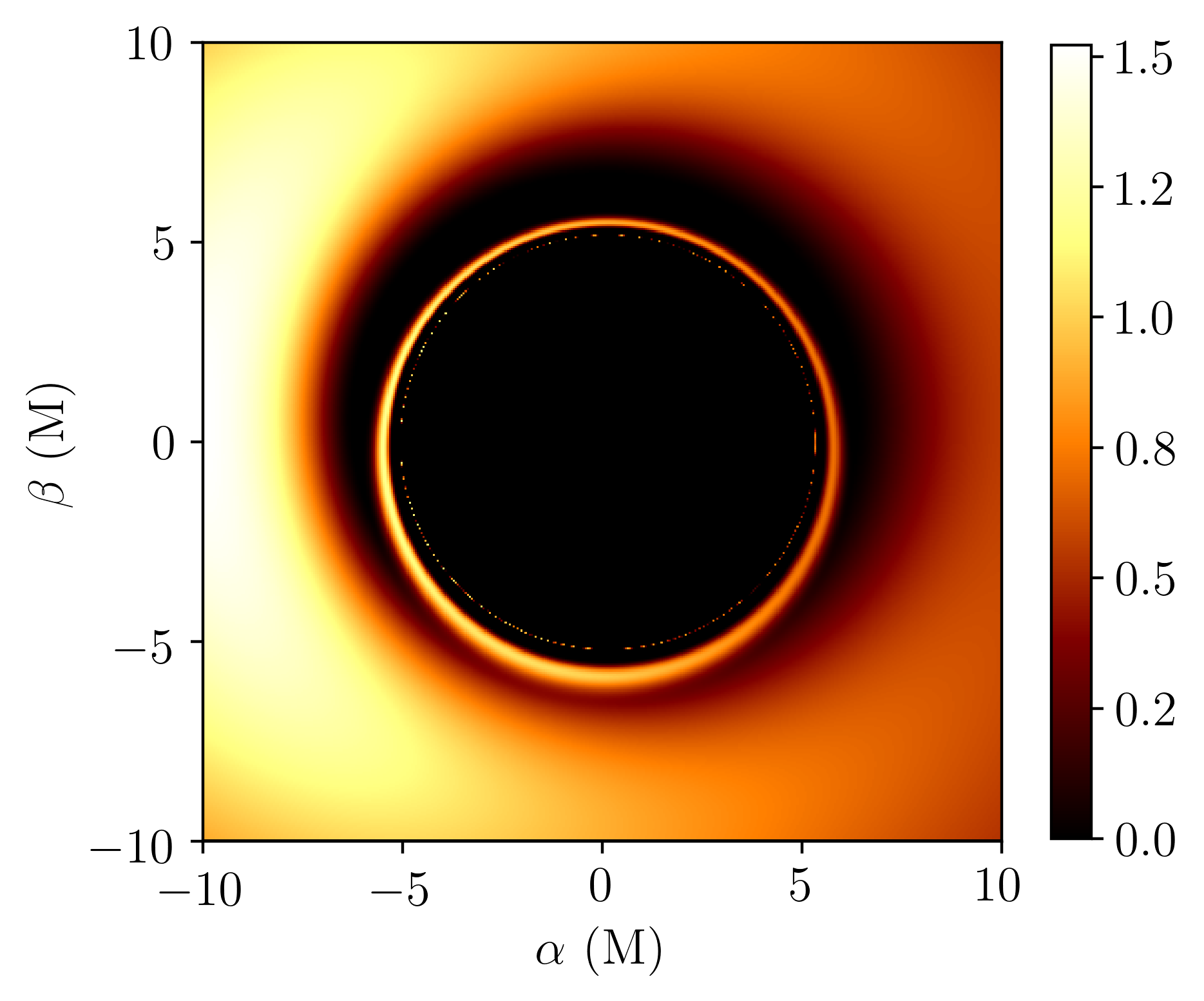}
    \includegraphics[scale = 0.55]{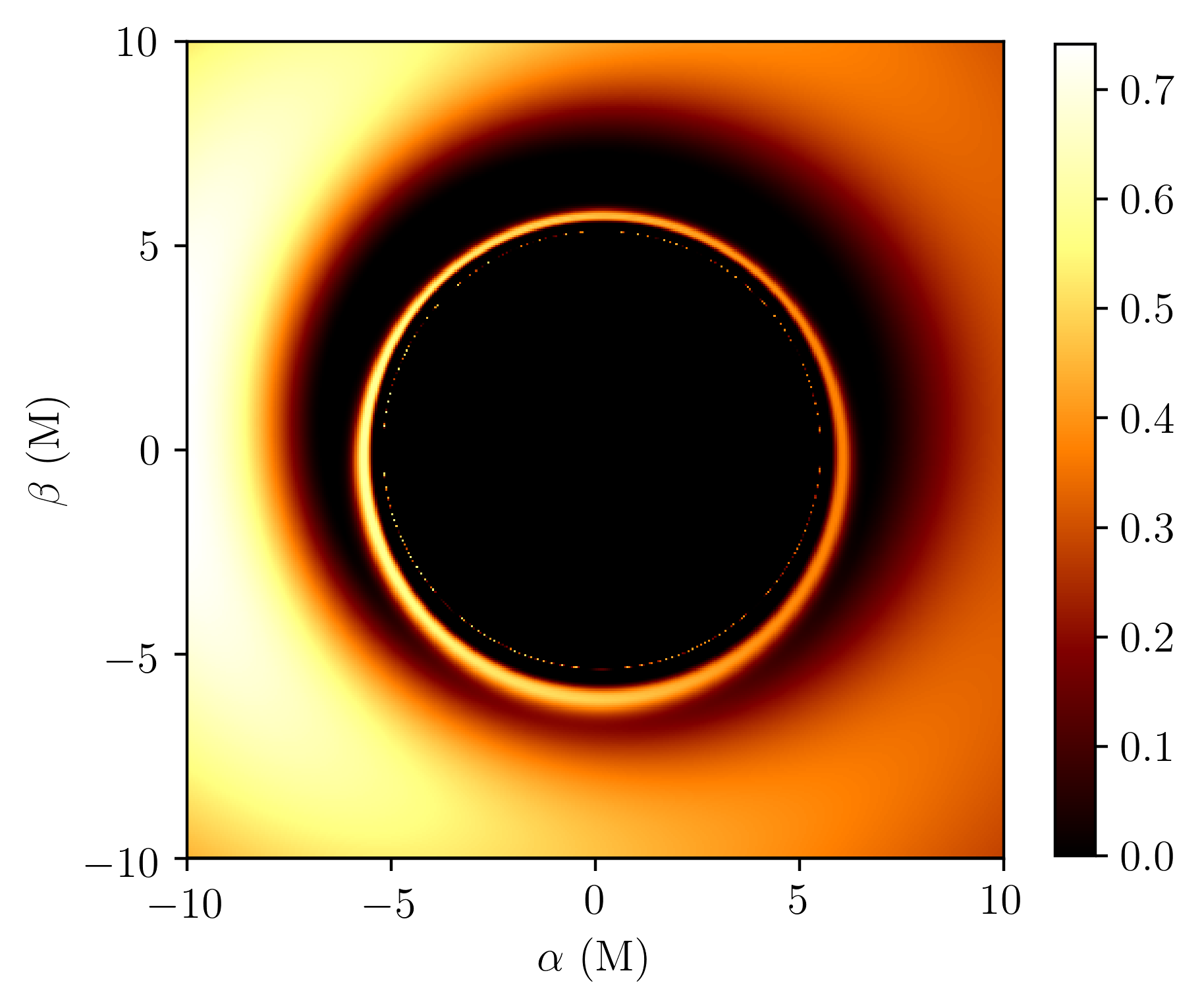}
    \includegraphics[scale = 0.55]{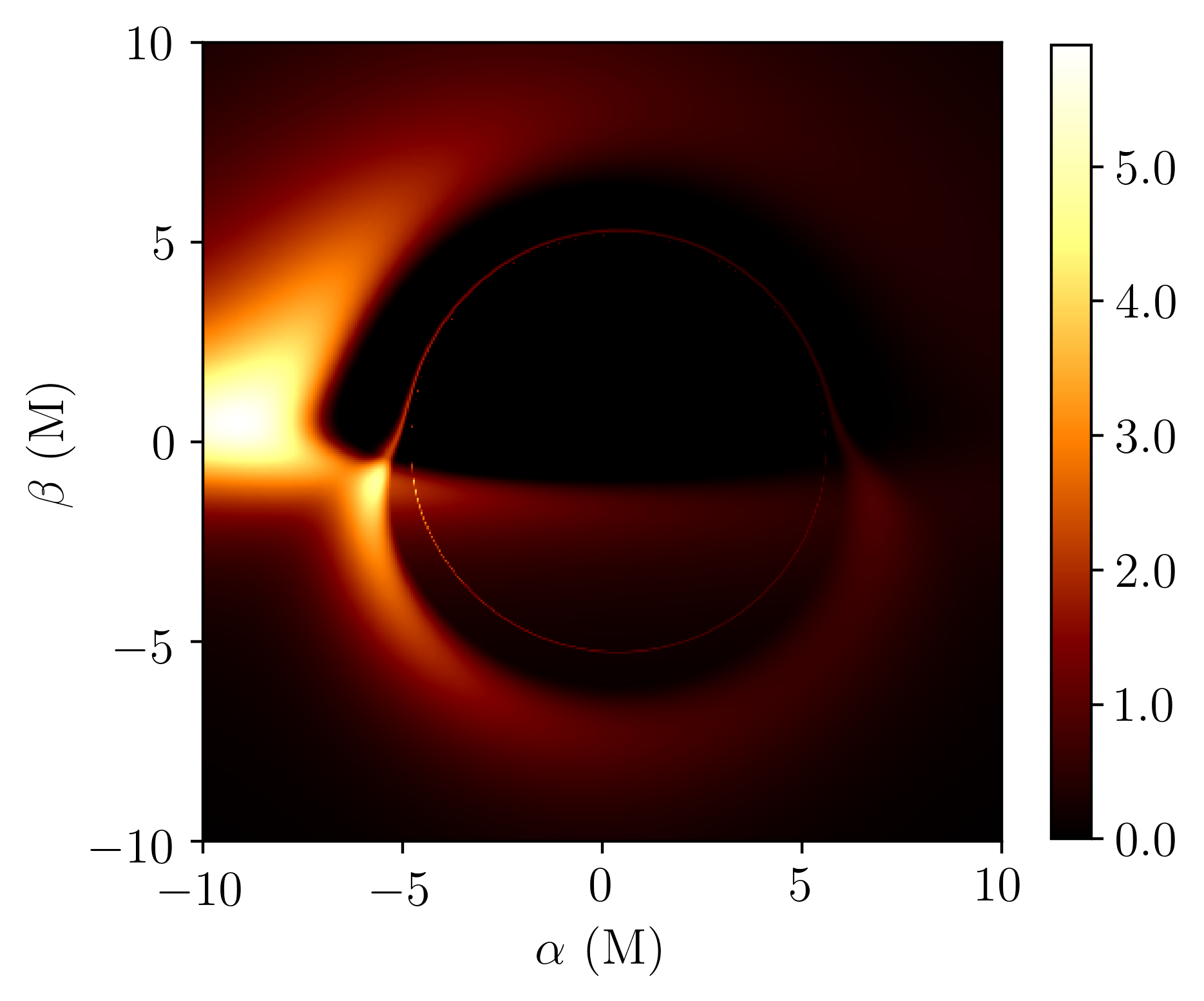}
    \includegraphics[scale = 0.55]{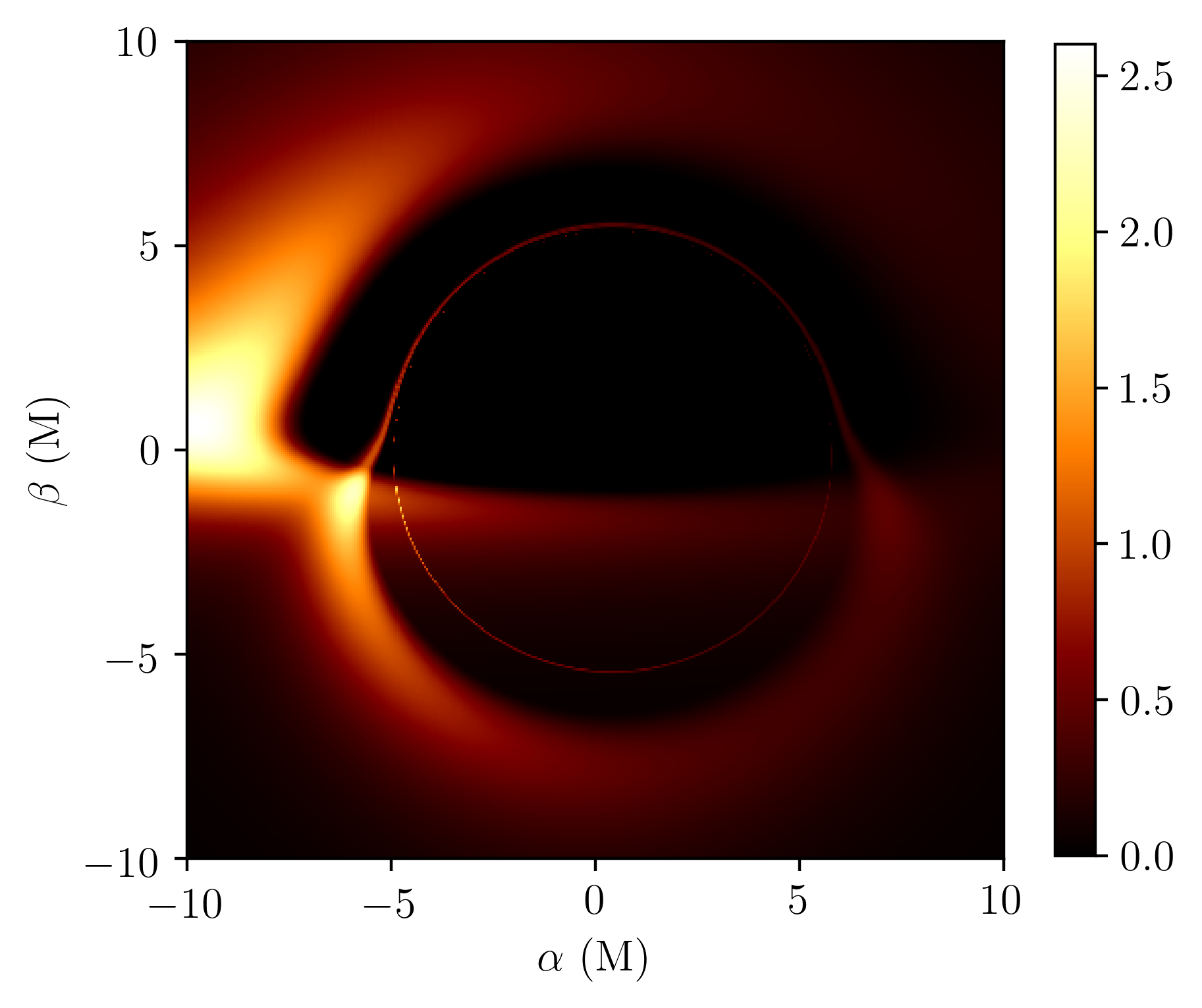}
    \includegraphics[scale = 0.55]{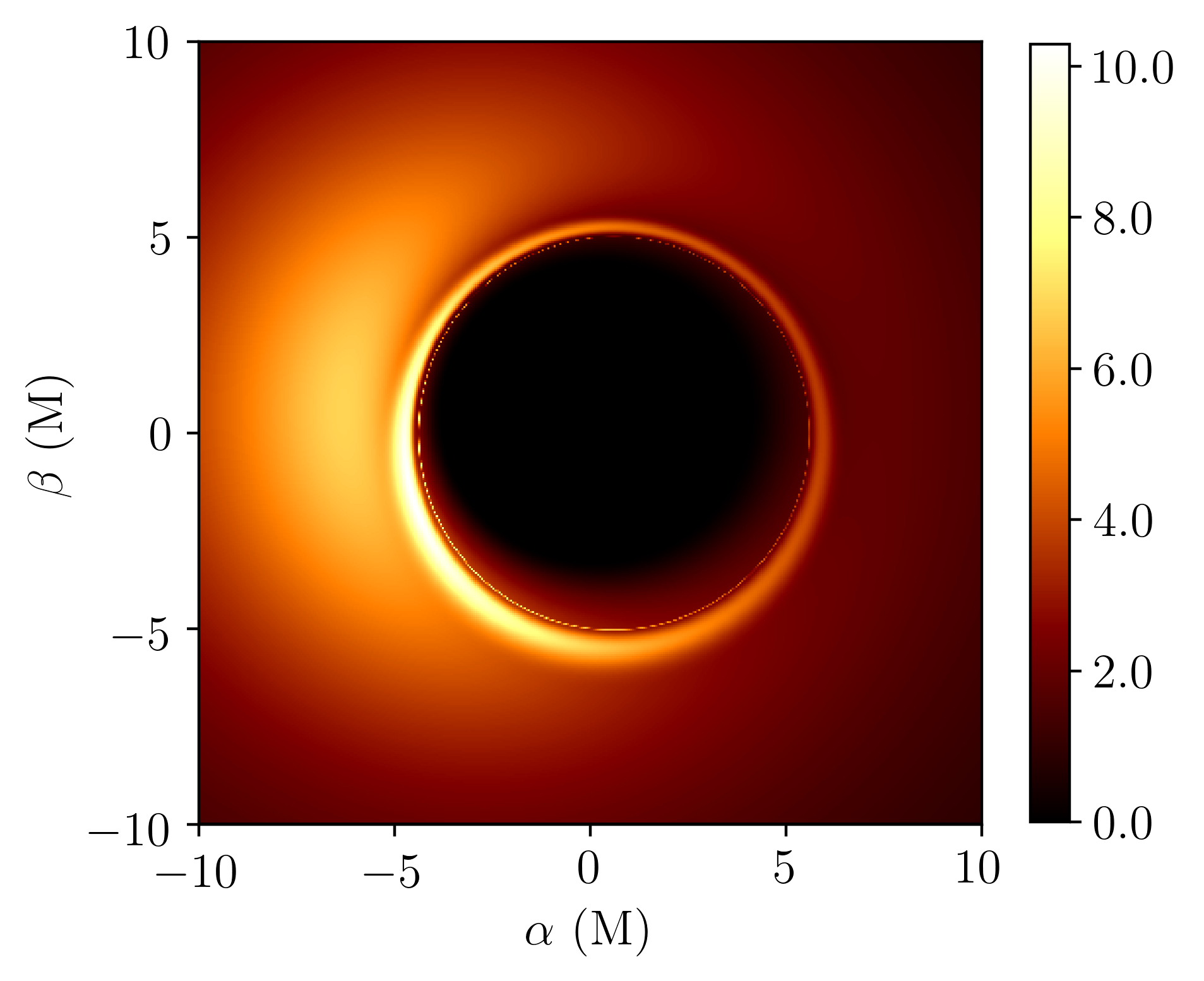}
    \includegraphics[scale = 0.55]{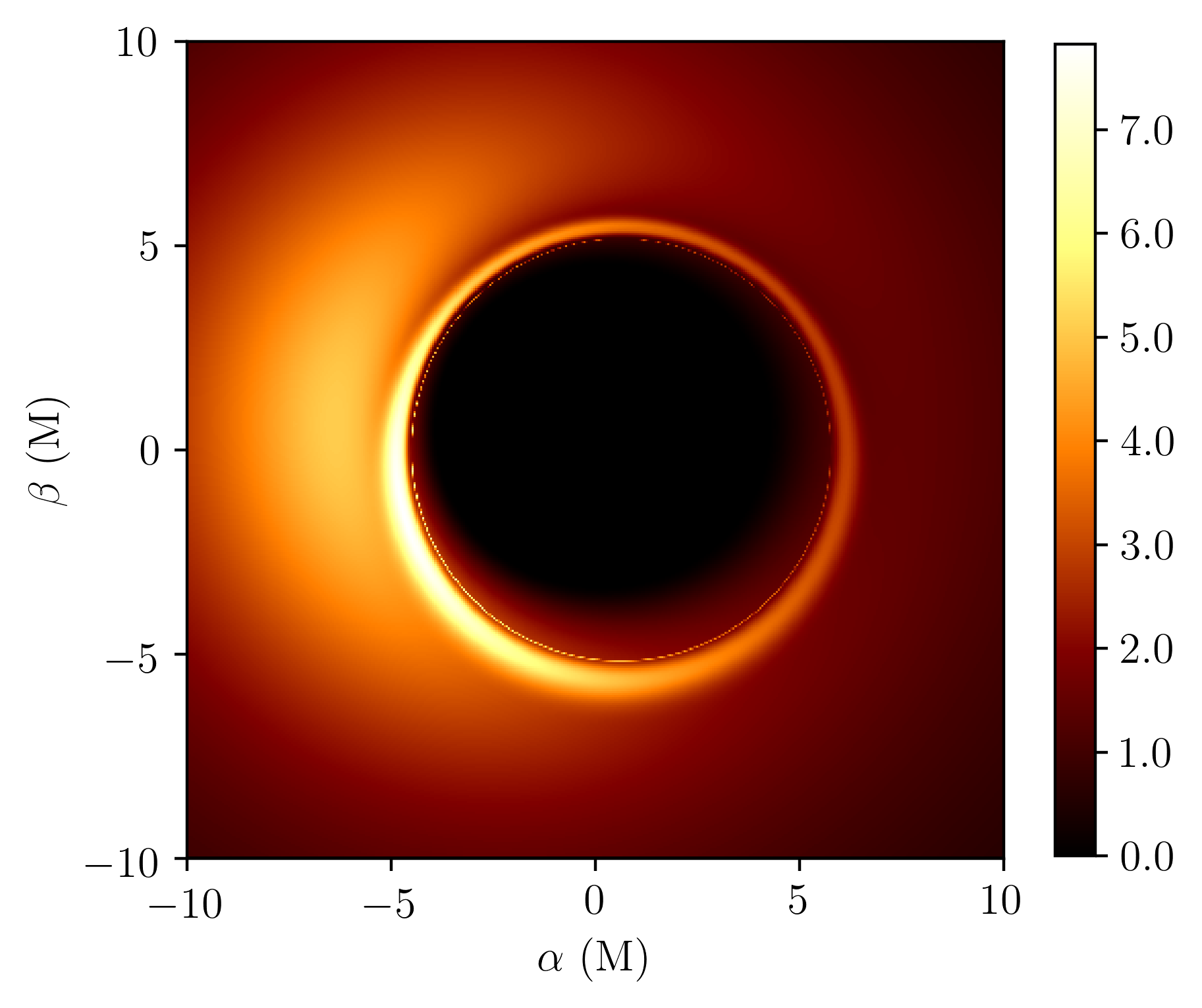}
    \includegraphics[scale = 0.55]{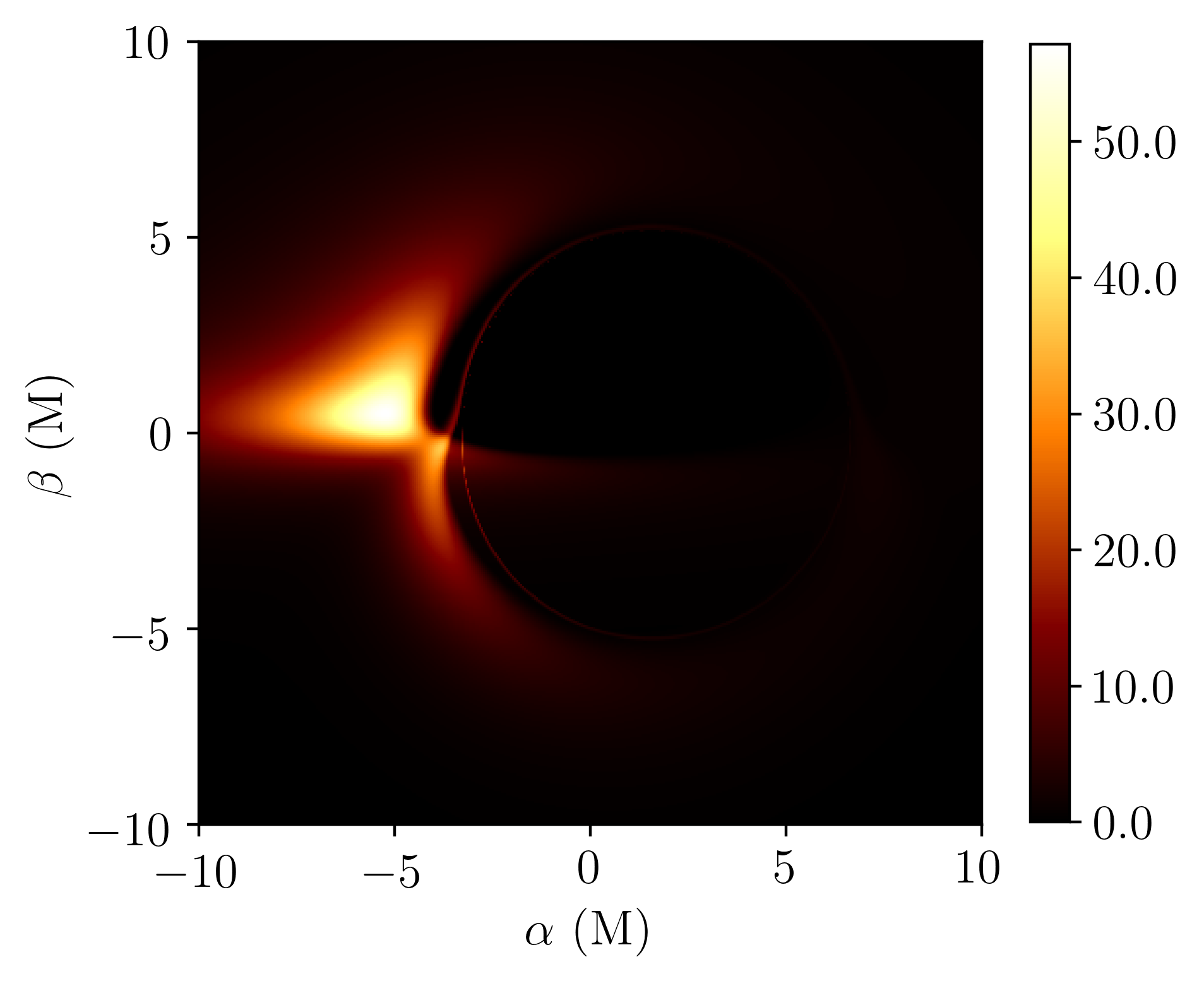}
    \includegraphics[scale = 0.55]{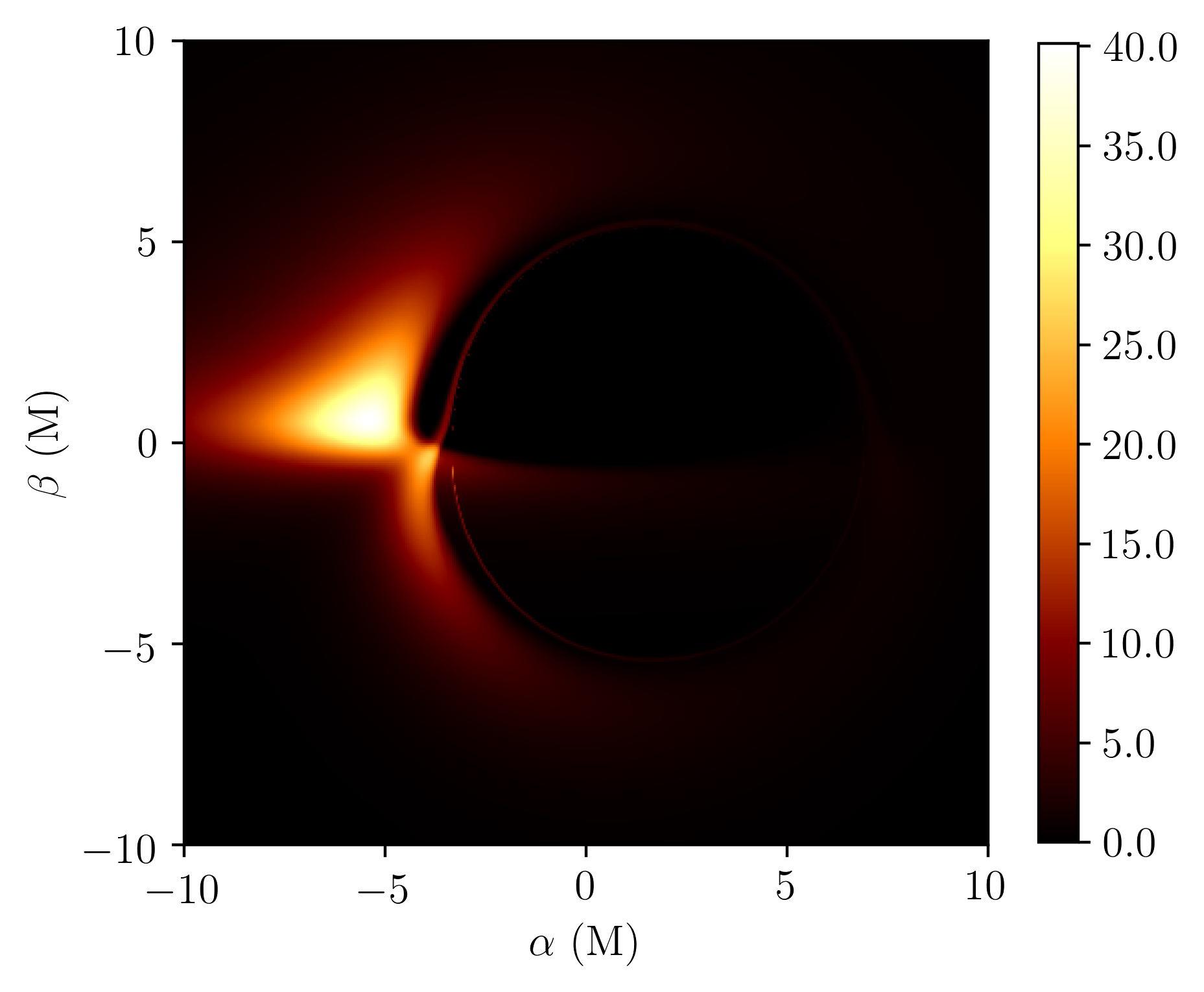}
\caption{The left column displays images of Kerr black holes, and the right column shows that of the dark matter black holes with the halo parameter set $\alpha,\beta,\gamma = (1,3,1)$. We also fix the halo parameter $\rho_sM^2=2\times10^{-5}$ and $r_s/M=30$. The first to fourth rows match four groups of parameter combinations which are $(a=0.2,\,i=20^\circ)$, $(a=0.2,\,i=80^\circ)$, $(a=0.8,\,i=20^\circ)$, and $(a=0.8,\,i=80^\circ)$ respectively. Color bars attached to each subplot describe the brightness of the black hole images. Brighter regions correspond to larger bolometric flux values. The source flux profile $\mathcal{F}(r)$ used to produce the observed images originates from Fig.~\ref{flux}. We set the accretion rate to $\dot{m}=1$ and multiply all flux values by a factor of $10^5$.}
\label{bhimage}
\end{figure*}

Fig.~\ref{bhimage} presents a comparison of bolometric images for standard Kerr black holes and rotating black holes with dark matter halos, with the halo parameter set $\alpha,\beta,\gamma=(1,3,1)$. Since $\gamma=1$ has more pronounced effect than $\gamma=0.5$, for simplicity, we only show one parameter set here. Throughout all calculations, we fix $\rho_sM^2=2\times10^{-5}$ and $r_s/M=30$. The four rows of subplots correspond to four combinations of black hole spin and viewing inclination angle: $(a=0.2,i=20^\circ)$, $(a=0.2,i=80^\circ)$, $(a=0.8,i=20^\circ)$, and $(a=0.8,i=80^\circ)$. The left column contains images of Kerr black holes, while the right column displays the counterparts of rotating black holes with dark matter halos. Each subplot is equipped with an independent color bar to quantify the observed flux magnitude, where larger color bar values correspond to higher radiative brightness. The radial flux profile of accretion disk $\mathcal{F}(r)$ adopted to generate these observational images is derived in Eq.~\eqref{eq:36} and shown in Fig.~\ref{flux}. We set the dimensionless accretion rate to $\dot{m}=1$ and multiply all flux values by a factor of $10^5$ for better visual presentation.

Horizontal comparisons between the left and right subplots under identical physical conditions reveal consistent and universal distinctions. For rotating black holes with dark matter halos, the black hole shadow, photon ring and Doppler-boosted bright spot all exhibit prominent outward expansion, the radial width of luminous photon rings becomes wider, and the entire imaging structure is less compact. In addition, the range of color bars clearly demonstrates that the peak radiative flux of dark matter black holes is always lower than that of Kerr black holes under identical parameters, which indicates that the spacetime correction induced by dark matter halo remarkably suppress the maximum radiative luminosity of accretion flows.

Black hole spin and viewing inclination modulate the distinguishability of imaging signatures arising from dark matter. At low inclination $i=20^\circ$, the images of both black hole models maintain nearly circular symmetric morphologies. Dark matter uniformly expands the overall size of photon rings and reduces the global radiative brightness, leading to relatively weak morphological discrepancies between the two models. As the viewing inclination rises to $i=80^\circ$, the image compress along the negative $\beta$ direction and forms an asymmetric teardrop shape, accompanied by a pronounced Doppler-enhanced bright spot on the left side of the black hole. The Doppler bright spot of Kerr black holes possesses a higher radiative peak, whereas dark matter black holes feature lower radiative peaks. From the perspective of radial distribution, the dark matter halos produce stronger expansion and luminosity suppression on outer-field photons.

\section{Conclusion}\label{sec6}

This paper systematically investigates the spacetime of rotating black holes embedded in dark matter halos, and quantitatively analyzes the effects of dark matter distribution on black hole geometry, thin accretion disk radiation, and observational features of black hole imaging, through comparisons with vacuum Kerr black holes. 

Starting from the corresponding static black hole with Zhao's general profile, we employed the Newman--Janis algorithm to construct its rotating counterpart and obtained a Kerr-like spacetime in which all dark matter effects are encoded in the modified radial dependent mass function $m(r)$. This framework allows us to systematically study how different dark matter distributions affect the near-horizon geometry and the observable properties of rotating black holes. 

We then derive the core physical quantities of timelike and null geodesics in this spacetime. It is found that dark matter halos shift the ISCO radius and photon shell, as well as event horizon outward. Such modulation effects of dark matter halo become more prominent for black holes with lower spins. 

Results on thin accretion disk demonstrate that dark matter halo suppresses the peak radiative flux, alters the radial position of flux maxima, and reduces disk temperature, accompanied by modified differential and spectral luminosity distributions. In the outer disk region, the influences of black hole spin and dark matter halo decay rapidly, where the effects of spin and dark matter halo become negligible. 

Ray-tracing imaging further reveals observational discrepancies. For the rotating black hole with dark matter halos, the critical curves and emission rings exhibit prominent outward expansion. Dark matter halos cause the black hole Doppler-boosted bright spot to expand outward, with an extended overall image structure. Compared with Kerr black holes, dark matter black holes exhibit systematically lower image luminosity, indicating that dark matter halo-induced spacetime corrections suppress the luminosity of accretion flows and increase the gravitational redshifts.

This study provides potential disk luminosity and imaging criteria to observationally discriminate rotating black holes with dark matter halos from standard Kerr black holes. Future work can test the relevant theoretical predictions using observational data from the Event Horizon Telescope, and extend to sophisticated accretion disk models that better mimic realistic astrophysical environments, to advance our understanding of dark matter effects in strong-gravity regimes.

\section*{Acknowledgments}
This work is supported by the National Natural Science Foundation of China (Grant No.~12505073). Zhen Li acknowledges the financial support from the Start-up Funds for Doctoral Talents at Jiangsu University of Science and Technology.
\\
\\


\begin{thebibliography}{99}

\bibitem{EHT2019M87}
K.~Akiyama et al. (Event Horizon Telescope),
Astrophys.~J.~Lett., \textbf{875}, L1 (2019).
\bibitem{EHT2022SgrA}
K.~Akiyama et al. (Event Horizon Telescope),
Astrophys.~J.~Lett., \textbf{930}, L17 (2022).
\bibitem{Kang2025}
J.-L. Kang \emph{et al.}, Mon. Not. R. Astron. Soc. \textbf{538}, 121 (2025).
\bibitem{Casura2024}
S. Casura, D. Ili\'{c}, J. Targaczewski, N. Raki\'{c}, and J. Liske, Mon. Not. R. Astron. Soc. \textbf{534}, 182 (2024).
\bibitem{Arevalo-Gonzalez2025}
F. Arevalo-Gonzalez \emph{et al.}, Mon. Not. R. Astron. Soc. \textbf{544}, 2737 (2025).
\bibitem{Juodzbalis2026}
I. Juod\v{z}balis \emph{et al.}, Mon. Not. R. Astron. Soc. \textbf{546}, 1 (2026).

\bibitem{Tremou2026}
E. Tremou \emph{et al.}, Mon. Not. R. Astron. Soc. \textbf{546}, 1 (2026).
\bibitem{John2024}
C. John, K. De, M. Lucchini, E. Behar, E. Kara, M. MacLeod, C. Panagiotou, and J. Wang, Mon. Not. R. Astron. Soc. \textbf{535}, 2633 (2024).
\bibitem{Tetarenko2016}
B. E. Tetarenko, G. R. Sivakoff, C. O. Heinke, and J. C. Gladstone, Astrophys. J. Suppl. Ser. \textbf{222}, 15 (2016).

\bibitem{Buisson2019}
D. J. K. Buisson \emph{et al.}, Mon. Not. R. Astron. Soc. \textbf{490}, 1350 (2019).
\bibitem{Novikov1973}
I.~D. Novikov, K.~S. Thorne, Black Holes (Les Astres Occlus), New York, 343 (1973).
\bibitem{Page1974}
D.~N. Page, K.~S. Thorne, Astrophys. J. \textbf{191}, 499 (1974).
\bibitem{Shakura1973}
N.~I. Shakura and R.~A. Sunyaev, Astron. Astrophys. \textbf{24}, 337 (1973).

\bibitem{Hankla2025}
A. M. Hankla, J. Dexter, and N. Scepi, Mon. Not. R. Astron. Soc. \textbf{541}, 3184 (2025).
\bibitem{Hagen2024}
S. Hagen, C. Done, J. D. Silverman, J. Li, T. Liu, W. Ren, J. Buchner, A. Merloni, T. Nagao, and M. Salvato, Mon. Not. R. Astron. Soc. \textbf{534}, 2803 (2024).
\bibitem{Luminet1979}
J.-P. Luminet, Astron. Astrophys. \textbf{75}, 228 (1979).

\bibitem{Hu2025}
S. Hu, D. Li, and C. Deng, J. Cosmol. Astropart. Phys. \textbf{2025}, 036 (2025).
\bibitem{Kumar2020}
R. Kumar and S. G. Ghosh, J. Cosmol. Astropart. Phys. \textbf{2020}, 053 (2020).
\bibitem{Li2021}
G.-P. Li and K.-J. He, J. Cosmol. Astropart. Phys. \textbf{2021}, 037 (2021).
\bibitem{Chael2021}
A. Chael, M. D. Johnson, and A. Lupsasca, Astrophys. J. \textbf{918}, 6 (2021).
\bibitem{Santibanez2025}
M. J. Santib\'{a}\~{n}ez-Armenta, G. Magallanes-Guij\'{o}n, S. Mendoza, and A. Cruz-Osorio, Mon. Not. R. Astron. Soc.: Lett. \textbf{540}, L54 (2025).
\bibitem{Chang2025}
C.-J. Chang and J.-F. Kiang, Mon. Not. R. Astron. Soc. \textbf{544}, 2599 (2025).

\bibitem{Hou2022}
Y. Hou, Z. Zhang, H. Yan, M. Guo, and B. Chen, Phys. Rev. D \textbf{106}, 064058 (2022).
\bibitem{Feng2024}
H. Feng, R.-J. Yang, and W.-Q. Chen, Astropart. Phys. \textbf{166}, 103075 (2025).
\bibitem{Guerrero2021}
M. Guerrero, G. J. Olmo, D. Rubiera-Garcia, and D. S\'{a}ez-Chill\'{o}n G\'{o}mez, J. Cosmol. Astropart. Phys. \textbf{2021}, 036 (2021).
\bibitem{Liu2022}
C. Liu, S. Yang, Q. Wu, and T. Zhu, J. Cosmol. Astropart. Phys. \textbf{2022}, 034 (2022).
\bibitem{BisnovatyiKogan2022}
G. S. Bisnovatyi-Kogan and O. Yu. Tsupko, Phys. Rev. D \textbf{105}, 064040 (2022).
\bibitem{Afrin2023}
M. Afrin and S. G. Ghosh, Mon. Not. R. Astron. Soc. \textbf{524}, 3683 (2023).

\bibitem{guo1}
Y. Hou, Z. Zhang, H. Yan, M. Guo, B. Chen, Phys. Rev. D \textbf{106}, 064058 (2022)
\bibitem{guo2}
J. Peng, M. Guo, X.H. Feng, Chin. Phys. C., \textbf{45}, 085103 (2021)
\bibitem{guo3}
Z. Zhang, Y. Hou, M. Guo, B. Chen, JCAP, \textbf{05}, 032 (2024)
\bibitem{yong1}
T.~Y.~Chen, Y.~Z.~Li and X.~M.~Kuang, Annals Phys. \textbf{488}, 170408 (2026).
\bibitem{yong2}
Y.~Meng, X.~J.~Wang, Y.~Z.~Li and X.~M.~Kuang,Eur. Phys. J. C., \textbf{85}, 627 (2025).
\bibitem{wang}
Z.~L.~Wang, Phys. Rev. D \textbf{112}, 6 (2025).

\bibitem{Kerr1963}
R. P. Kerr, Phys. Rev. Lett. \textbf{11}, 237 (1963).

\bibitem{Johnson1}
M. D. Johnson \emph{et al.}, Sci. Adv. \textbf{6}, eaaz1310 (2020).

\bibitem{Gralla1}
S. E. Gralla, D. E. Holz, and R. M. Wald, Phys. Rev. D \textbf{100}, 024018 (2019).
\bibitem{deGraaff2024}
A. de Graaff, A. Pillepich, and H.-W. Rix, Astrophys. J. Lett. \textbf{967}, L40 (2024).
\bibitem{Shankar2025}
R. Shankar, A. Prakash, and A. Mehta, Mon. Not. R. Astron. Soc. \textbf{540}, 2269 (2025).
\bibitem{Sarkar2026}
S. Sarkar, P. Biswas, V. Kalinova, N. Roy, N. N. Patra, and S. Kurapati, Mon. Not. R. Astron. Soc. \textbf{546}, 1 (2026).

\bibitem{Liu2023}
D. Liu, Y. Yang, A. \"{O}vg\"{u}n, Z.-W. Long, and Z. Xu, Eur. Phys. J. C \textbf{83}, 565 (2023).

\bibitem{Hou2018}
X. Hou, Z. Xu, M. Zhou, and J. Wang, J. Cosmol. Astropart. Phys. \textbf{07}, 015 (2018).

\bibitem{Shen2024}
Z. Shen, A. Wang, Y. Gong, and S. Yin, Phys. Lett. B \textbf{855}, 138797 (2024).

\bibitem{Mora2025}
N. U. Mora, H. Chaudhary, S. Capozziello, F. Atamurotov, G. Mustafa, and U. Debnath, Phys. Dark Universe \textbf{47}, 101804 (2025).

\bibitem{Zhu2019}
K. Jusufi, M. Jamil, P. Salucci, T. Zhu, and S. Haroon, Phys. Rev. D \textbf{100}, 044012 (2019).

\bibitem{37} 
H. S. Zhao, MNRAS, \textbf{278}, 488 (1996).

\bibitem{nfw}
J.~F.~Navarro, C.~S.~Frenk and S.~D.~M.~White, Astrophys. J., 462, 563-575 (1996).
\bibitem{heq}
L. Hernquist, Astrophys. J., 356, 359 (1990).
\bibitem{prof1}
H.~C.~Plummer, Mon. Not. R. Astron. Soc. 71, 460 (1911).
\bibitem{prof2}
H.~Dejonghe, MNRAS, 224, 13 (1987).
\bibitem{prof3}
T.~de Zeeuw, MNRAS, 216, 273 (1985).
\bibitem{prof4}
H.~J.~Rood, T.~L.~Page, E.~C.~Kintner and I.~R.~King, Astrophys. J. 175, 627 (1972).
\bibitem{prof5}
W.~Jaffe, MNRAS, 202, 995 (1983).
\bibitem{prof7}
M.~Baes and H.~Dejonghe, A\&A, 393, 485 (2002).
\bibitem{prof8}
B.~Moore, T.~Quinn, F.~Governato, J.~Stadel and G.~Lake, MNRAS, 310, 1147 (1999).
\bibitem{prof9}
W.~Dehnen, MNRAS, 265, 250 (1993).

\bibitem{zli}
Z.~Li, Astron. Astrophys. 710, A336 (2026).

\bibitem{nj1}
E. T. Newman and A. I. Janis, J. Math. Phys., \textbf{6}, 915 (1965).
\bibitem{nj2}
E. T. Newman, E. Couch, K. Chinnapared, et al., J.
Math. Phys., \textbf{6}, 918 (1965).
\bibitem{nj3}
M.~Azreg-A{\"\i}nou, Phys. Rev. D, 90, 064041 (2014).
\bibitem{nj4}
M.~Azreg-A{\"\i}nou, Eur. Phys. J. C, 74, 2865 (2014).
\bibitem{nj5}
H.~Erbin, Universe, \textbf{3},  19 (2017).
\bibitem{st}
A.~Burinskii, E.~Elizalde, S.~R.~Hildebrandt and G.~Magli, Phys. Rev. D, 65, 064039 (2002).

\bibitem{Bardeen1972}
J.~M. Bardeen, W.~H. Press, and S.~A. Teukolsky, Astrophys. J. \textbf{178}, 347 (1972).

\bibitem{Chen2025}
H.~Chen, W.~Fan, and X.~Y. Chew, Eur. Phys. J. C \textbf{85}, 338 (2025).

\bibitem{l1}
K. Boshkayev, A. Idrissov, O. Luongo, D. Malafarina, Mon. Not. R. Astron. Soc. {\bf496}, 1115 (2020).
\bibitem{l2}
E. Kurmanov, K. Boshkayev, R. Giambò, T. Konysbayev, O.
Luongo, D. Malafarina, H. Quevedo, Astrophys. J. {\bf925}, 210 (2022)
\bibitem{l3}
K. Boshkayev, T. Konysbayev, Y. Kurmanov, O. Luongo, D. Mala-farina, Astrophys. J. {\bf936}, 96 (2022)
\bibitem{l4}
Y. Kurmanov, K. Boshkayev, T. Konysbayev, M. Muccino, O. Luongo, A. Urazalina, A. Dalelkhankyzy, F. Belissarova, M.
Alimkulova, Phys. Dark Universe {\bf48}, 101917 (2025)
\bibitem{Li2025}
Z.~Li and X.-K.~Guo, Eur. Phys. J. C \textbf{85}, 679 (2025).

\bibitem{Cunningham1973} 
C. T. Cunningham and J. M. Bardeen, Astrophys. J. \textbf{183}, 237 (1973).

\bibitem{Gralla:2020a}
S. E. Gralla and A. Lupsasca, Phys.\ Rev.\ D {\bf 101}, 044031 (2020).
\bibitem{Gralla:2020b}
S. E. Gralla, A. Lupsasca and D. P. Marrone, Phys.\ Rev.\ D {\bf 102}, 124004 (2020).


\bibitem{Cardenas2023}
A.~C\'ardenas-Avenda\~no, A.~Lupsasca, and H.~Zhu, Phys.~Rev.~D \textbf{107}, 043030 (2023).

\end{thebibliography}
\end{document}